\documentclass[aps,prl,twocolumn,superscriptaddress,longbibliography,floatfix]{revtex4-2}
\usepackage{amsmath,amssymb,graphicx,bm,xcolor,dsfont}
\usepackage[colorlinks=true,linkcolor=blue,citecolor=blue,urlcolor=blue]{hyperref}
\newcommand{\Tr}{\mathrm{Tr}}
\newcommand{\one}{\mathds{1}}
\newcommand{\X}{\mathcal{X}}
\newcommand{\Xb}{\bar{\mathcal{X}}}
\newcommand{\ii}{\mathrm{i}}
\newcommand{\ODM}{\mathcal O_{\rm DM}}

\begin{document}

\title{Dzyaloshinskii--Moriya interactions and deconfined criticality in \texorpdfstring{SrCu$_2$(BO$_3$)$_2$}{SrCu2(BO3)2}}

\author{Yasir Iqbal}
\email{yiqbal@physics.iitm.ac.in}
\affiliation{Department of Physics, Indian Institute of Technology Madras, Chennai 600036, India}

\begin{abstract}
SrCu$_2$(BO$_3$)$_2$ is the leading candidate for deconfined criticality,
yet theories of its critical regime omit the Dzyaloshinskii--Moriya (DM)
interactions, which gap the kagome Dirac spin liquid. Adding them to the
SU(2) gauge theory, we show that time reversal and a spin--orbit-locked
glide forbid any gauge-singlet spinon mass to all orders. In-plane DM only
twists the spinons, reconstructing the nodes below $0.03$--$0.17$\,K, while
out-of-plane DM is irrelevant and SO(5) anisotropy arises only at second
order. The material thus remains a platform for deconfined criticality,
which NMR can test, since a $[100]$ field induces a N\'eel moment and a
$[001]$ field does not.
\end{abstract}

\maketitle

SrCu$_2$(BO$_3$)$_2$, a Shastry--Sutherland (SS) antiferromagnet~\cite{ShastrySutherland,Kageyama1999,MiyaharaUeda1999,MiyaharaUeda2003},
is the most studied material candidate~\cite{LYSV2019} for deconfined quantum criticality~\cite{Senthil2004a,Senthil2004b}. Under pressure its plaquette-singlet
phase gives way to an antiferromagnet~\cite{Zayed2017,Guo2020,LarreaJimenez2021}, and whether a deconfined
critical point controls that transition is debated. Field-tuned $^{11}$B
NMR finds first-order signatures that weaken with pressure and suggests a
proximate, likely multicritical, deconfined critical point, with
quantum-critical scaling at $2.4$\,GPa near the transition field
$H_c\simeq5.7$\,T~\cite{Cui2023}. Zero-field
calorimetry finds a first-order transition with a bicritical
point~\cite{Guo2025}. In applied fields near the critical pressure a
$T$-linear specific heat of open origin appears, possibly from
fractionalized spin excitations~\cite{Guo2026,Song2026}. Between the SS model's plaquette and N\'eel phases,
numerical studies find a narrow gapless spin-liquid phase~\cite{Yang2022,Viteritti2025,Corboz2026} (a $\mathbb Z_2$ Dirac spin liquid~\cite{Maity2025}), a weakly first-order transition~\cite{Xi2023,Qian-2025}, or a
continuous one with emergent symmetry~\cite{Wang2024,LYSV2019}. On the
theory side, an SU(2) gauge theory of Dirac
fermions and adjoint Higgs fields, introduced for the square lattice~\cite{ShackletonThomsonSachdev2021}, unifies the SS Dirac spin liquids with
the N\'eel and valence-bond orders and supports an SO(5)-symmetric
pseudocritical regime~\cite{Feuerpfeil2026,Maity2025,Maity2026}.

These field theories retain the SU(2) spin symmetry. The material,
however, has Dzyaloshinskii--Moriya (DM) couplings of up to a few tenths
of a meV~\cite{Cepas2001,Kodama2005,Mazurenko2008,Romhanyi2015}. These
have so far been included only in the dimer phase, at the
dimer--plaquette transition~\cite{LarreaJimenez2021} and in spin-wave
analyses of the N\'eel phase~\cite{Fogh2024PRL,Cederholm2026}, and their absence from
the theory of the deconfined scenario has been noted~\cite{Cui2024two}.
On the kagome lattice DM destabilizes the Dirac spin liquid: the
out-of-plane DM component couples linearly to a time-reversal-even spin-triplet fermion
mass of quantum-spin-Hall type, the hexagon vector chirality. It is relevant
and was argued to drive magnetic order~\cite{Hermele2008}, as found
numerically above $D\simeq0.1J$~\cite{Cepas2008} and in variational
studies~\cite{Kagome2407,Ferrari-2023}. The DM-driven transition has also
been analyzed as one out of a $\mathbb Z_2$ spin liquid of bosonic
spinons~\cite{Huh2010}. Whether spin--orbit coupling can
similarly gap the SS Dirac spin liquid, or otherwise destroy its
criticality, has not been addressed.

\begin{figure*}[t]
\includegraphics[width=\textwidth]{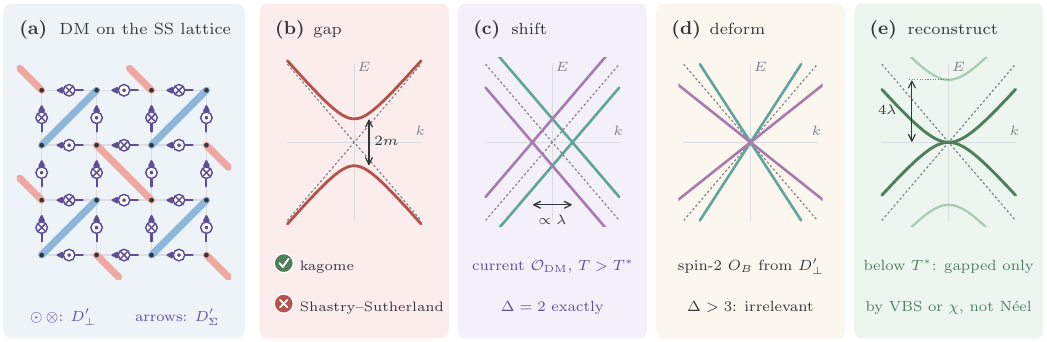}
\caption{Allowed and forbidden effects of DM on the spinon Dirac cones.
(a) SS lattice with the $[110]$ (blue) and $[1\bar10]$ (rose) dimers and
the inter-dimer DM vectors for bonds $i\to j$ along $+x$ and $+y$, drawn
for $D'_\perp,D'_\Sigma>0$. Shown are $D'_\perp$ ($\odot$/$\otimes$) and the in-plane component along each bond's tetragonal axis, which sets $D'_\Sigma$; the other in-plane component, which does not enter $\lambda$, and the intra-dimer $\bm D$ are omitted. (b)--(e) Band cuts through a Dirac
node, with the unperturbed cone dashed. (b) A mass opens a gap. DM does
this on kagome~\cite{Hermele2008} but not on the SS lattice, where every
gauge-singlet mass is forbidden (Table~\ref{tab:masses}). (c) Among
gauge-singlet bilinears the in-plane couplings induce only $\lambda\ODM$,
a symmetry-allowed combination of components of the conserved spin-current multiplet that acts as a spin vector potential and shifts
the spin-resolved cones (teal, plum) by an amount proportional to
$\lambda$. (d) At linear order $D'_\perp$ induces only the spin-2 operator
$O_B$, which splits the cone velocities and is irrelevant at a unitary
critical point. (e) Below $T^*=4\lambda$, near the critical point, the cones reconstruct into a
quadratic band touching, with $E=\pm[\sqrt{v^2k^2+4\lambda^2}\pm2\lambda]$
for Dirac velocity $v$. Its only gauge-singlet gap channels are the
valence-bond doublet and the scalar chirality $\chi$.}
\label{fig:ops}
\end{figure*}

\begin{figure}[t]
\includegraphics[width=\linewidth]{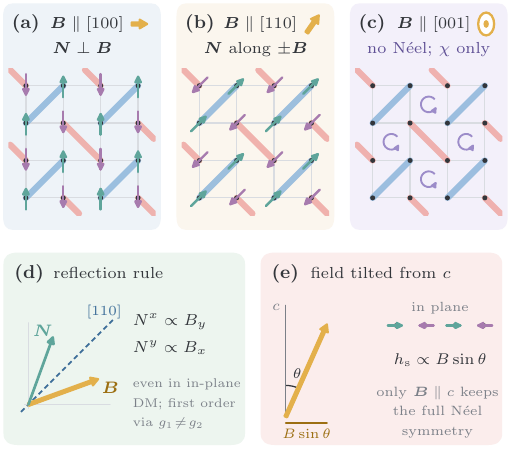}
\caption{Field selection rule for the staggered moment
$\bm N=\sum_i\epsilon_i\bm S_i$ ($\epsilon_i=\pm1$ on the $[110]$ (blue)
and $[1\bar10]$ (rose) dimers, spins drawn teal and plum), for a positive
coefficient, whose sign is not universal. (a) $\bm B\parallel[100]$
induces $\bm N\parallel[010]$, perpendicular to $\bm B$. (b)
$\bm B\parallel[110]$ induces $\bm N$ along $[110]$. (c)
$\bm B\parallel[001]$ induces no N\'eel moment at any order; among the
masses it couples only to the scalar chirality. (d) The induced $\bm N$ is, up to that
sign, the mirror image of $\bm B$ in the $[110]$ dimer axis, $N^x\propto B_y$, $N^y\propto B_x$, even in
the in-plane DM couplings and first order in a dimer-dependent in-plane
$g$ tensor. (e) The in-plane part of a field tilted by $\theta$ from $c$
is conjugate to the in-plane N\'eel order parameter, through the
in-plane staggered field $h_s$.}
\label{fig:field}
\end{figure}

Here, we answer this question by classifying the fermion bilinears of
the continuum theory under the spin--orbit-locked symmetry group, in which
every lattice operation also rotates the spins as an axial vector. We
obtain the following results, summarized in Figs.~\ref{fig:ops}
and~\ref{fig:field}. (i) Time reversal and a
single locked glide forbid all six gauge-singlet fermion masses. (ii) Among the gauge-singlet
bilinears, the in-plane DM couplings enter the critical theory only as a
conserved spin current, with dimension exactly 2 and renormalization-group
eigenvalue exactly 1 at a spin-rotation-invariant fixed point. Below the resulting scale $T^*$, near the critical
point and in a two-dimensional sample, the Dirac cones reconstruct into a quadratic band touching that,
among gauge-singlet orders, only valence-bond or chirality order can gap. (iii) The largest inter-dimer
coupling, the out-of-plane $D'_\perp$, induces no fermion bilinear without
gradients and, with one gradient, only a spin-2 operator of dimension
above 3 at any unitary critical point. At second order DM induces an
SO(5)-breaking tensor whose relevance is disputed. (iv) A field along
$[100]$ induces a staggered N\'eel moment perpendicular to it, a field
along $[110]$ one along the field, and a field along $[001]$ none.

\emph{Model and locked symmetry.}---We study
\begin{equation}
H=J\!\sum_{\rm dimers}\!\bm S_i\!\cdot\!\bm S_j
+J'\!\sum_{\rm squares}\!\bm S_i\!\cdot\!\bm S_j
+\sum_{\langle ij\rangle}\bm D_{ij}\!\cdot\!(\bm S_i\times\bm S_j),
\label{eq:H}
\end{equation}
with the DM vectors allowed by the crystal structure. Above
$T_s\simeq395$\,K the space group is $I4/mcm$ and only the alternating
out-of-plane inter-dimer component $D'_\perp$ is allowed~\cite{Sparta2001,Cepas2001}; below $T_s$ the buckling of the CuBO$_3$
layers ($I\bar42m$, tetragonal up to the monoclinic onset at
$4$--$5$\,GPa~\cite{Loa2005,Zayed2014}) adds an in-plane intra-dimer $D$ and two in-plane
inter-dimer components $D'_{\parallel,s}$, $D'_{\parallel,ns}$~\cite{Sparta2001,Kodama2005,Cheng2007,Romhanyi2015}. With DM the
spins can no longer be rotated independently of the lattice. A lattice
operation $O$ acts on a spin as the axial vector
$\bm S\to\det(O)\,O\,\bm S$, and the textures of
Fig.~\ref{fig:ops}(a) lie in the invariant subspace of this action on bond
vectors~\cite{companion}. In the buckled structure the glide $G_x$ and the
fourfold rotation survive only combined with the horizontal mirror, so
they act with the spin rotations $C_{2x}$ and $R_z(-\pi/2)$,
\begin{equation}
\hat G_x=G_x\otimes C_{2x},\ \
\hat S_4=C_4\otimes R_z(-\tfrac{\pi}{2}),\ \
\hat\sigma_{xy}=\sigma_{xy}\otimes C_2^{[1\bar10]},
\label{eq:locked}
\end{equation}
together with $\hat\sigma_{x\bar y}$ and time reversal $\mathcal T$.

The low-energy theory is the SU(2) gauge theory of
Refs.~\cite{Feuerpfeil2026,ThomsonSachdev}: eight Majorana-matrix Dirac
fields $\X$ at the $\pi$-flux point, with valley matrices $\mu^a$ and
spin matrices $\sigma^a$, coupled to an SU(2) gauge field and to three
adjoint Higgs fields $\Phi_{1,2,3}$, of which $\Phi_1$ and $\Phi_2$ are
critical at the SO(5) point. Condensing $\Phi_1,\Phi_2$ yields the gapless
$\mathbb Z_2$ state Z3000, condensing $\Phi_3$ the U(1) state U800 (labels of
Ref.~\cite{Maity2025}). We project the lattice symmetry action, including the
projective gauge transformations, onto the Dirac subspace and let each
generator act on the spin index through its locked rotation. The Higgs
fields are spin singlets, so their potential is unchanged.

\emph{No symmetry-allowed gauge-singlet fermion mass at any order in spin--orbit coupling.}---The gauge-singlet fermion bilinears
$\Tr[\Xb M\X\sigma^b]$ at zeroth gradient order span a 36-dimensional
space. Six are masses~\cite{ThomsonSachdev}: the SO(5) vector $\Tr[\Xb\Gamma^j\X]$, made of the
two valence-bond-solid (VBS) components $V_{x,y}$ and the three N\'eel
components $N^a$, and the flux singlet $\Tr[\Xb\X]$, which carries the
scalar spin chirality $\chi$~\cite{WangNahum2017}. The
other 30 are components of the
conserved SO(5)-adjoint currents. The ten SO(5)-adjoint scalars vanish identically by the Majorana reality condition of the SU(2)
theory~\cite{companion}. Table~\ref{tab:masses}
lists the action of $\mathcal T$ and $\hat G_x$ on the six masses. The
N\'eel and chirality masses are odd under time reversal, and the two VBS
masses, the only $\mathcal T$-even ones, are both odd under the locked
glide, so no linear combination is invariant. The locked space group
alone suffices: $\hat G_x$ together with $\hat\sigma_{xy}$ leaves
no mass invariant, and time reversal is an additional reason for the
N\'eel and chirality masses~\cite{SM}. Any perturbation that
preserves time reversal and the crystal symmetry, DM at any order or
the symmetric exchange anisotropies, can therefore generate no
gauge-singlet fermion mass. The same holds without spin--orbit
coupling, where the unlocked glide plays the role of $\hat G_x$. Thus, locking the spins to the lattice allows no new gauge-singlet mass.

\emph{Role of the SU(2) gauge structure.}---Spin--orbit coupling preserves
time reversal and breaks spin rotations, so at linear order it can induce
only time-reversal-even operators that carry spin. On kagome one
such operator is a mass, the quantum-spin-Hall-type triplet~\cite{Hermele2008}. In the SU(2)
gauge theory of the SS spin liquid that whole family, the SO(5)-adjoint
scalars, vanishes identically. The 36-dimensional operator space leaves no other candidate. Every time-reversal-even, spin-carrying gauge-singlet bilinear without
gradients is a component of a conserved current, and the only spin-carrying masses, the N\'eel
components, are odd under time reversal. The remaining masses are the
order parameters of competing states, each of which breaks time reversal
or the lattice symmetry, and locking makes none of them symmetric. The two cases thus differ in their gauge group. In the U(1) theory of the
kagome Dirac spin liquid the quantum-spin-Hall masses are gauge
invariant, whereas in the SU(2) theory of the $\pi$-flux state they
vanish as gauge singlets. In Z3000 their gauge-adjoint partners reappear,
$\propto D\langle\Phi\rangle$, but leave the nodes gapless for material
couplings (End Matter).

\begin{table}[b]
\caption{Parities of the six Dirac masses under time reversal and the
locked glide $\hat G_x=G_x\otimes C_{2x}$~\cite{companion}. Every mass is
odd under at least one of the two, so no mass is invariant.}
\label{tab:masses}
\begin{ruledtabular}
\begin{tabular}{lccc}
mass & operator & $\mathcal T$ & $\hat G_x$\\
\hline
VBS $V_x$, $V_y$ & $\Tr[\Xb\mu^{z}\X]$, $\Tr[\Xb\mu^{x}\X]$ & $+$ & $\bm-$\\
N\'eel $N^{x,y,z}$ & $\Tr[\Xb\mu^y\X\sigma^a]$ & $\bm-$ & $-,+,+$\\
chirality $\chi$ & $\Tr[\Xb\X]$ & $\bm-$ & $-$\\
\end{tabular}
\end{ruledtabular}
\end{table}

The theorem concerns explicit local masses in the symmetric state and
requires the full layer group. A shear strain along $[110]$ lowers it and
allows a VBS mass, and a magnetic field allows the masses of
Fig.~\ref{fig:field}. The theorem does not exclude
spontaneous symmetry breaking, pairwise annihilation of displaced
Dirac nodes of opposite chiral winding (excluded numerically for
material couplings in Z3000, End Matter), or confinement by monopoles in the U(1) descendant U800.
Its single symmetry-allowed monopole is the SO(5) singlet with or
without DM~\cite{companion}.

\begin{table}[t]
\caption{Perturbations of the SS Dirac spin liquid and the operators they
induce. ``Exact'' marks statements fixed by symmetry, current conservation
or unitarity ($^*$for a unitary critical point). The dimension of the
$\mathbf{14}$ (``open'') is disputed between the large-$N$ gauge theory and the
fuzzy-sphere SO(5) model, and the signs of its DM-induced coefficients are
mean-field results.}
\label{tab:summary}
\begin{ruledtabular}
\begin{tabular}{lll}
perturbation & induced operator & status\\
\hline
any $\mathcal T$-even, symmetric & no gauge-singlet mass & exact\\
in-plane DM & $\ODM$, $\Delta=2$ & exact\\
$D'_\perp$, no gradient & no bilinear & exact\\
$D'_\perp$, one gradient & spin-2 $O_B$, $\Delta>3$ & exact$^*$\\
$(D'_\perp)^2$, $\lambda^2$ & $\mathbf{14}$, $\Delta=3.46$ or $1.456$ & open\\
$\bm B\parallel c$ & $\chi$ only, no N\'eel & exact\\
$\bm B\perp c$ & $(N^x,N^y)\propto(B_y,B_x)$ & exact\\
\end{tabular}
\end{ruledtabular}
\end{table}

\emph{Linear order and the band touching.}---Of the 36 bilinears, DM can induce only those invariant under the
symmetries its couplings preserve; they are summarized in
Table~\ref{tab:summary}~\cite{companion}. With $D'_\perp$ alone the model keeps the spin rotations
about $c$ and the high-temperature symmetries, and no gauge-singlet bilinear
without gradients is invariant, even without $\mathcal T$. With the in-plane couplings a single
bilinear is induced,
\begin{equation}
\delta\mathcal L=\lambda\,\ODM,\qquad
\ODM=J^x_x-J^y_y,
\label{eq:ODM}
\end{equation}
where $J^a_\mu$ is the current of spin component $a$ along $\mu$
($x\parallel[100]$, $y\parallel[010]$); $\ODM$ is the spin--orbit-locked component
of the SU(2) spin currents allowed by the $\bar42m$ point group
[Fig.~\ref{fig:ops}(c)]. Time reversal removes two further current
combinations. Because $J^a_\mu$ are Noether currents of the spin-rotation-invariant
fixed point, every component of the current multiplet, and hence $\ODM$, has the protected
scaling dimension $\Delta=2$ at that fixed point, and $\lambda$ has renormalization-group eigenvalue exactly 1.
A constant $\ODM$ is a non-abelian spin vector potential,
$A_x\propto S^x$, $A_y\propto-S^y$. A position-dependent spin rotation maps it at linear order onto spin-dependent node displacements, but because $[A_x,A_y]\neq0$ it cannot be removed, and near the critical point, where the Higgs splitting is small, it reconstructs the Dirac nodes below the scale $T^*=4\lambda$ [Fig.~\ref{fig:ops}(e)]. The eightfold node splits
into branches at $\pm4\lambda$ and a $d$-wave quadratic band touching,
$H\propto(v^2/4\lambda)[(k_x^2-k_y^2)\mathcal A_1+2k_xk_y\mathcal A_2]$
with anticommuting matrices $\mathcal A_{1,2}$, protected by a chiral symmetry. Projected onto it, the N\'eel masses
$N^{x,y}$ vanish and $N^z$ only splits the bands, whereas $V_{x,y}$ and
$\chi$ open full gaps. Below $T^*$, N\'eel bilinear order therefore cannot gap the spinons of the band touching. A quadratic band touching is unstable to weak interactions~\cite{Sun2009,VafekYang2010}. Here, among gauge-singlet orders, only valence-bond order and a chiral spin liquid can gap it, and a one-loop flow places the instability at $T^{**}\sim(0.03$--$0.1)\lambda$, in the millikelvin range (End Matter).

A lattice Wick decoupling on the $\pi$-flux state estimates the
coefficient at mean field, $\lambda=\chi_0D'_\Sigma/(2\sqrt2)$, with
$\chi_0=0.2395$ the nearest-neighbor singlet amplitude and
$D'_\Sigma=D'_{\parallel,s}+D'_{\parallel,ns}$, $\sqrt2$ times the
in-plane inter-dimer component along the tetragonal axis of each bond,
so that $T^*\simeq\sqrt2\chi_0|D'_\Sigma|\simeq0.34|D'_\Sigma|$~\cite{companion}. The intra-dimer $D\simeq0.22$--$0.35$\,meV~\cite{Kodama2005,Mazurenko2008}, the largest DM coupling, does not enter
at this order, and $D'_\perp$ cancels. Since $D$ transforms like the
in-plane inter-dimer couplings, it can renormalize $\lambda$ beyond mean
field, a contribution we do not estimate. Spectroscopy constrains only the
combination $D'_{\parallel,s}\mp(J'/2J)D$~\cite{Cheng2007,Romhanyi2015},
so $D'_\Sigma$ rests on the one available density-functional calculation~\cite{Mazurenko2008}, which gives $T^*\simeq0.03$--$0.17$\,K, uncertainties included (End Matter). The mean-field $T^*$ lies below the temperatures of
about $1.2$--$1.6$\,K at which the pressurized material orders in zero
field~\cite{Guo2025}. The thermodynamic crossover lies lower still, near
$(0.1$--$0.15)T^*$ (End Matter).

The out-of-plane $D'_\perp\simeq0.09$--$0.18$\,meV~\cite{Cepas2001,Romhanyi2015,Mazurenko2008}, the largest inter-dimer coupling, enters at linear order only
at first gradient order, through the $(xy)$ component of a flavored spin-2 current,
\begin{equation}
O_B=\Tr\bigl[\Xb\,\mu^y\sigma^z(\gamma^y\ii\overset{\leftrightarrow}{D}_x
+\gamma^x\ii\overset{\leftrightarrow}{D}_y)\X\bigr],
\label{eq:OB}
\end{equation}
which splits the cone velocities without gapping them
[Fig.~\ref{fig:ops}(d)]. A non-conserved spin-2 primary of a unitary
conformal field theory has dimension $\Delta>3$~\cite{Minwalla1998,Evans1967}, so $O_B$ is irrelevant
at the SO(5) fixed point if that point is unitary; at leading order in $1/N$, with $N=12$, its
dimension is $3+16/(5\pi^2)\simeq3.32$ (End Matter).

\emph{Second order and SO(5) breaking.}---At second order DM induces the rank-2 symmetric tensor of SO(5), the $\mathbf{14}$
($\mathbf5\otimes\mathbf5=\mathbf1\oplus\mathbf{10}\oplus\mathbf{14}$):
$D'_\perp$ through $O_B\times O_B$, since $O_B$ carries the N\'eel$_z$
flavor $\mu^y\sigma^z$, and the in-plane couplings at order $\lambda^2$.
Its SO(3)$\times$SO(2)-invariant component is the N\'eel--VBS
anisotropy, the tuning operator of the QCD$_3$ and sigma-model
descriptions~\cite{ThomsonSachdev,WangNahum2017,Zhou2024fuzzy}; for it
DM only shifts the critical pressure. The rest are spin anisotropies that break the emergent
symmetry, easy-axis from $D'_\perp$ and easy-plane from $\lambda$ at mean
field (End Matter). Whether they matter depends on the dimension of the
$\mathbf{14}$, on which the two descriptions of the SO(5) point
disagree: it is relevant in the fuzzy-sphere study of the SO(5) sigma
model ($\Delta=1.456$), obtained there in the walking region~\cite{Zhou2024fuzzy}, and irrelevant at leading order ($\Delta\simeq3.46$)
in the large-$N$ Higgs theory, tuned instead by the SO(5)-singlet Higgs
mass~\cite{companion}. If the $\mathbf{14}$ is irrelevant, no DM-induced SO(5)-breaking four-fermion operator acts above $T^*$ within the $O(1/N)$ four-fermion spectrum, whose mixing with gauge and Higgs composites outside the $\mathbf{14}$ is not computed~\cite{companion}. If it
is relevant, second-order anisotropies do act there, at scales estimated
from below $0.1$\,K to about $1$\,K~\cite{companion}. In either case the
DM-induced Yukawa couplings remain subleading (End Matter). Neither a DM
mass nor $\lambda$ at linear order can produce the first-order signatures
and the bicritical point near $1.3$\,K~\cite{Guo2025}. They may instead
come from a second-order anisotropy, the interlayer coupling or
disorder~\cite{companion}.

\emph{Field selection rule.}---A magnetic field lowers the symmetry to
the stabilizer of $(\text{DM},\bm B)$: the locked elements that fix
$\bm B$ and the products of $\mathcal T$ with those that reverse it.
Classifying the 36 bilinears under these groups gives the activated
masses of Fig.~\ref{fig:field}~\cite{companion}. For $\bm B\parallel[001]$
the only activated mass is the flux singlet, so the field couples linearly
to the scalar chirality. For the Zeeman coupling this requires
spin--orbit locking, because
without it the magnetization is a spin triplet and $\chi$ a singlet. No
N\'eel component is allowed at any order in $B$. For an in-plane field the
N\'eel moment is activated at linear order with the tensor
$N^x\propto B_y$, $N^y\propto B_x$. Up to a non-universal sign, $\bm N$
is the mirror image of $\bm B$ in the $[110]$ dimer axis, perpendicular to
$\bm B\parallel[100]$ and on the axis of $\bm B\parallel[110]$. The order in the DM couplings follows from the
horizontal mirror $\hat\sigma_h$, which acts on the spins as a global
$C_{2z}$ rotation. It maps the model onto itself with every in-plane DM
component and the in-plane components of $\bm B$ and $\bm N$ reversed,
so the in-plane N\'eel response to an in-plane field is even in the
in-plane DM couplings and absent at first order. The intra-dimer $D$ alone
already produces it at second order, and $D'_\perp$ alone, which conserves
$S^z$, gives none at any order. A dependence of the in-plane
$g$ tensor on the dimer orientation, allowed by the Cu site symmetry,
enters at first order in spin--orbit coupling. With components $g_1$
along the dimer and $g_2$ across it, the Zeeman term contains the
staggered field $\tfrac12(g_1-g_2)\mu_B\sum_i\epsilon_i(B_yS_i^x+B_xS_i^y)$
with the same tensor, which together with DM produces the field-induced gap in
Cu benzoate~\cite{OshikawaAffleck1997}. It induces a moment wherever the
staggered
susceptibility $\partial\langle\bm N\rangle/\partial h_s$ is finite,
which excludes only the exact dimer singlet at zero temperature. The two inversion twins of the crystal~\cite{Sparta2001} carry opposite in-plane DM, so the even response has
the same sign in both.

Because the rule rests only on the lattice symmetry, it also holds in the dimer phase, where it is distinct from the first-order intra-dimer staggered moments of Refs.~\cite{Kodama2005,Miyahara2004,Miyahara2007}, which are antiparallel within each dimer and lie along $c$ for an in-plane field. The rule predicts that a $^{11}$B NMR comparison of $\bm B\parallel[100]$ with
$\bm B\parallel[001]$ at matched Zeeman energy under pressure and below $H_c$, above the
plaquette ordering temperature, shows a N\'eel-pattern splitting, with
the moments of each dimer parallel, only for the in-plane field, provided
the moment transfers a hyperfine field to the boron site. Because a $c$-axis field also splits the lines, through the intra-dimer
staggered moments, the two field directions are distinguished by the
pattern of the splitting. No experiment has made this comparison (End Matter). In the spin-liquid regime the $[100]$ moment grows toward the critical pressure as $m_s\propto B\,\xi^{3-2\Delta_N}$ ($\xi$ the correlation length, $\Delta_N$ the N\'eel dimension), with exponent $0.08$ at leading order in $1/N$, small enough for a regular background to dominate, and of order one for $\Delta_N\simeq0.63$~\cite{Nahum2015}. For field-tuned tests of deconfined criticality, any in-plane component of the field, as in a tilted field, is linearly conjugate to the in-plane N\'eel order parameter [Fig.~\ref{fig:field}(e)] and rounds a critical point on a scale of $0.1$--$1$\,K for the $8.6^\circ$ tilt of Ref.~\cite{Cui2023} and an illustrative $g$ anisotropy (End Matter).

\emph{Outlook.}---The spin--orbit coupling of SrCu$_2$(BO$_3$)$_2$
cannot give its Dirac spinons a gauge-singlet mass. At linear order it only
twists the spinons through a conserved current, which below $T^*$
reconstructs the nodes into a band touching that, among gauge-singlet orders, is open to valence-bond and
chiral order but not to N\'eel order. Above $T^*$ it can act, apart from Yukawa couplings that start below $2\%$ of the one that bounds the pseudocritical window, only through second-order rank-2 anisotropies. Their relevance, fixed by the
dimension of the SO(5) $\mathbf{14}$, and the signs and sizes of
their DM-seeded coefficients remain open. The no-mass theorem uses
only the time-reversal oddness of the N\'eel and chirality masses, the
lattice parities of the spin-singlet VBS masses, which locking cannot
change, and the absence of quantum-spin-Hall masses in the SU(2) theory.
It therefore carries over to any lattice realization of this gauge theory
whose SO(5) vector consists of N\'eel and VBS order parameters and whose crystal symmetry leaves no VBS combination invariant.

\emph{Acknowledgments.}---
The work of Y.I. was performed, in part, at the Aspen Center for Physics, which is supported by a grant from the Simons Foundation (1161654,
Troyer). This research was also supported in part by grant NSF PHY-2309135 to the Kavli Institute for
Theoretical Physics (KITP). Y.I. acknowledges support from the Abdus Salam International Centre for Theoretical Physics (ICTP) through the Associates Programme, from the Simons Foundation through Grant No. 284558FY19, from IIT Madras through the Institute
of Eminence (IoE) program for establishing QuCenDiEM (Project No. SP22231244CPETWOQCDHOC), and the International Centre for Theoretical Sciences (ICTS), Bengaluru for participation in the Discussion Meeting ``Fractionalized Quantum Matter'' (code: ICTS/DMFQM2025/07).
The author acknowledges the use of the AI tools Claude Opus 5.5 and
Fable 5.1 (Anthropic) to assist in developing code, cross-checking some
of the calculations, and preparing parts of the manuscript. The author
reviewed all results and changes and takes full responsibility for the
content.

\emph{Data availability.}---The data that support the findings of this
article are available from the author upon reasonable request.

\section*{End Matter}

\emph{Continuum theory and locked action.}---Projecting the lattice projective symmetry group of Ref.~\cite{Feuerpfeil2026} onto the node subspace, with each generator carrying its axial spin rotation, gives the continuum action of the locked generators of Eq.~\eqref{eq:locked} and $\hat\sigma_{x\bar y}=\sigma_{x\bar y}\otimes C_2^{[110]}$ (Supplemental Material~\cite{SM}). The emergent SO(5) has generators $T^j=\{\mu^y,\sigma^a,\mu^x\sigma^a,\mu^z\sigma^a\}$ and vector $\Gamma^j=\{\mu^x,\mu^z,\mu^y\sigma^a\}$, with Dirac matrices $\gamma^0=\rho^y$, $\gamma^x=\ii\rho^z$, $\gamma^y=\ii\rho^x$.

\emph{The gauge-singlet bilinears.}---The Hermitian gauge-singlet bilinears at zeroth gradient order form the
gauge-invariant subspace of the 120 quadratic Majorana forms, of dimension
36: the six masses and the 30 components of the currents
$\Tr[\Xb\gamma^\mu T^j\X]$. The ten scalars $\Tr[\Xb T^j\X]$ vanish
identically, because the reality condition makes each of them the trace of
a product that changes sign under transposition~\cite{companion}. The
tables of all 36 operators under the unlocked, buckled-locked and
high-temperature-locked generators are given in the Supplemental Material~\cite{SM}. The gauge-adjoint bilinears transform as in
Ref.~\cite{Feuerpfeil2026}. Reference~\cite{ThomsonSachdev} agrees except for the time-reversal signs
of the N\'eel-flavored currents $\Tr[\sigma^a\Xb\mu^y\sigma^b\gamma^\mu\X]$,
which the time-reversal oddness of the N\'eel mass fixes as in Ref.~\cite{Feuerpfeil2026} (Supplemental Material~\cite{SM}). The no-mass theorem
uses only two columns of these tables
(Table~\ref{tab:masses}). The tabulated spin labels,
related to physical axes by $x\to-x$, leave these columns unchanged and
turn Eq.~\eqref{eq:ODM} into
$\Tr[\Xb\gamma^x\X\sigma^x]+\Tr[\Xb\gamma^y\X\sigma^y]$. A construction directly in
physical axes gives the same classification and invariant $J^x_x-J^y_y$~\cite{SM}.

At the critical point and in the Higgs phases the gauge-adjoint
bilinears enter through Yukawa couplings. With $D'_\perp$ no new Yukawa
coupling is allowed. With the in-plane components three appear, including
the first zeroth-order coupling of $\Phi_3$~\cite{companion}. At the
multicritical point the first two are relevant: one is an SO(5) and
Lorentz partner of the Yukawa coupling $y$ of Ref.~\cite{Feuerpfeil2026},
the other, a scalar, is less relevant than $y$ at $O(1/N)$. Only the intra-dimer
$D$ generates them, starting at $6\times10^{-4}$--$2\times10^{-2}$ of $y$ barring a cancellation among the Z3000 amplitudes,
so they remain subleading throughout the pseudocritical window~\cite{companion}. In Z3000
they become quadratic terms proportional to $D\langle\Phi\rangle$, which
the singlet argument does not control. There the nodes are protected by a
chiral symmetry: the Bogoliubov Hamiltonian with every DM term
anticommutes with $(-1)^{i_x+i_y}$ times an on-site rotation,
so each node
carries a winding number, and a direct diagonalization finds the nodes
gapless up to couplings far beyond the material values~\cite{companion}. In the U(1) state U800 the locked group admits a
zeroth-order $\Phi_3$ coupling only to a current but none to a mass~\cite{companion}. Non-magnetic quenched disorder couples a relevant random field to the valence-bond doublet and none to the chirality or N\'eel channels~\cite{companion}.

\emph{The coefficient $\lambda$.}---The Wick decoupling of $\bm D\cdot(\bm S_i\times\bm S_j)$ in the singlet
channel $\langle f^\dagger_{i\alpha}f_{j\beta}\rangle=\chi_{ij}
\delta_{\alpha\beta}$ gives the triplet hopping
$\tfrac{\ii}{2}\bm D\cdot(\chi_{ij}f^\dagger_j\bm\sigma f_i-{\rm h.c.})$.
On the $\pi$-flux state ($|\chi_{ij}|=\chi_0=0.23952$ in the thermodynamic
limit) the sum over the eight inter-dimer bonds of a cell projects onto
pure $\ODM$ with $\lambda=\chi_0D'_\Sigma/(2\sqrt2)$; $D'_\perp$ cancels, and the
intra-dimer $D$ drops out because the dimer amplitude vanishes on the
$\pi$-flux state. At linear order the DM term acts as a background spin gauge field. It
twists the spinon hopping by an angle of order $D'_\Sigma/J'$, so the node
splitting scales with the spinon velocity. The Fock value above and a
hopping $t\simeq0.4J'$ therefore give splittings that differ by a factor
of $1.7$, which sets the uncertainty of the mean-field prefactor.
The specific heat of the free spinons crosses over to the band-touching
regime only near $(0.1$--$0.15)T^*$, because the reconstructed band is
quadratic only well below $\lambda$~\cite{companion}.

Triplon spectroscopy fixes
$D'_{\parallel,s}\mp(J'/2J)D\simeq0.07$--$0.08$\,meV~\cite{Cheng2007,Romhanyi2015}
and is blind to $D'_{\parallel,ns}$ at linear order. The only tabulated
inter-dimer DM vectors~\cite{Mazurenko2008}, placed on their bonds with the
bond vectors of that calculation~\cite{MazurenkoPC}, give $D'_\perp\simeq-0.10$\,meV and an in-plane component
of $0.01$\,meV along the tetragonal axis of each bond. This
yields $|D'_\Sigma|\simeq0.014$\,meV, or $0.007$--$0.025$\,meV
once the rounding of the table and a rescaling to the
electron-spin-resonance value of $D'_\perp$ are included, and hence
$T^*\simeq0.03$--$0.1$\,K at the Fock level, rising to $0.17$\,K for
$t\simeq0.4J'$. These are ambient-pressure couplings; the buckling that allows the in-plane components may decrease under pressure~\cite{Loa2005}, which would lower $T^*$.

\emph{The band touching below $T^*$.}---At the node the Dirac Hamiltonian is
$h(\bm k)=v(k_x\rho^x-k_y\rho^z)+\lambda h_{\rm DM}$, where $h_{\rm DM}$ is
the single-particle matrix that $\ODM$ adds at $\bm k=0$. It splits the
eightfold multiplet into branches at $\pm4\lambda$ and four gapless modes. Second-order $k\cdot p$ theory gives
$H_{\rm QBT}=(v^2/4\lambda)[(k_x^2-k_y^2)\mathcal A_1+2k_xk_y\mathcal A_2]$
with $\{\mathcal A_1,\mathcal A_2\}=0$, Berry winding 2, isotropic at all
$k$. A $k^2\one$ term is forbidden by the chiral symmetry $\rho^y$, which
anticommutes with $h(\bm k)$ including the DM term. On the band-touching subspace, $N^x$ and $N^y$ vanish because
they live on the $\pm4\lambda$ branches, $N^z$ commutes with
$\mathcal A_{1,2}$ and splits the bands into pockets, and $V_x$, $V_y$ and
the flux singlet anticommute with both and open full gaps. On this
subspace the uniform magnetization obeys
$P(J^0\sigma^z)=-P(\text{flux})$, so the chirality carries a weak
$z$ moment and a $c$-axis field couples to it linearly. A one-loop flow of the 14 locked-invariant quartic couplings with the screened gauge field ends on a valence-bond ray for the material's exchange, and otherwise on a gapped chiral $\mathbb Z_2$ spin liquid, at $T^{**}\sim(0.03$--$0.1)\lambda$. This regime requires a two-dimensional sample and a pseudocritical window that reaches $T^*$~\cite{companion}.

\emph{The spin-2 operator and the second-order tensor.}---Being flavored, $O_B$ cannot mix with gauge or Higgs operators at $y=0$.
At $O(1/N)$, with $N=2N_f+4N_b=12$ the gauge-polarization count of two
Dirac valleys and two real adjoint Higgs triplets, the self-energy, vertex
and the two seagull graphs of the covariant derivative give
$\gamma_2=64C_g/(5\pi^2N)=16/(5\pi^2)\simeq0.32$ at Casimir $C_g=3$, independent of
the gauge parameter. At $C_g=1$ this reduces to the QED$_3$ flavor-adjoint
spin-2 anomalous dimension of Ref.~\cite{ZhouHe2023}, with $C_g$ placed as in
Ref.~\cite{Gracey1994}. The same rules give the valence-bond and N\'eel dimension
$\Delta_N=2-64C_g/(3\pi^2N)\simeq1.46$
(QED$_3$ adjoint value~\cite{RantnerWen2002,Hermele2005} at $C_g=1$)
and, at the $y=0$ large-$N$ Higgs fixed point, the scalar four-fermion
operator in the SO(5) $\mathbf{14}$ forms an isolated irrep with
$\Delta_{\mathbf{14}}=4-64/(\pi^2N)\simeq3.46$ for $N=12$ at this order. With
$64/\pi^2N\simeq0.54$ these values fix only signs and orderings, and whether the $\mathbf{14}$ is relevant is open. The Higgs sector is less controlled: its constraint field has $\Delta_\sigma=0.38$ at $O(1/N)$, below the unitarity bound $1/2$, so SO(5)-singlet dimensions at $N=12$ are not controlled~\cite{companion}.
The bound $\Delta>3$ for $O_B$ does not rest on the expansion; it assumes
a unitary critical point and does not apply in the walking scenario with
complex fixed points~\cite{WangNahum2017}. The seagull sign is fixed by gauge covariance, and the same rules give the full spin-$\ell$ tower~\cite{companion}.

At second order, a Kubo computation of the mass-channel responses with the
Wick-decoupled $D'_\perp$ operator gives rank-2 coefficients
$\tau_1=-0.18$ and $\tau_2=-1.94$ per $(D'_\perp/t)^2$, in the channel basis
$\tau_1\,{\rm diag}(3,3,-2,-2,-2)+\tau_2\,{\rm diag}(0,0,1,1,-2)$ on
$(V_x,V_y,N^x,N^y,N^z)$, nearly temperature independent and dominated by
the lattice scale. They describe an easy-axis anisotropy ($\tau_2<0$) with a N\'eel-side
tuning component ($\tau_1<0$)~\cite{companion}. The normalization is
arbitrary, so only signs and ratios are meaningful. The in-plane couplings give
$\tau_1,\tau_2>0$ of order $\lambda^2/T$, an easy-plane anisotropy with a
singlet-side tuning component. The $D'_\perp$ term dominates above
$T_\times\simeq0.01$--$0.3$\,K at the Fock level~\cite{companion}. The coefficients grow with the momentum cutoff, so
the sign is a lattice-scale, mean-field result; a classical bond-level
estimate of the same order, $D'^2_\perp/2J'^2$, is easy-plane instead.
The intra-dimer $D$, which has no mean-field image, the symmetric
exchange anisotropies and the dipolar coupling generate the same tensor at the same order and are not computed.
For a single bond, DM and the symmetric anisotropy of the same order are
equivalent to isotropic exchange by a local spin rotation~\cite{Shekhtman1992}.
If the $\mathbf{14}$ is relevant with $y_T=3-1.456\simeq1.54$, a
coupling $g_T$ becomes of order one at $E_T\simeq\Lambda g_T^{1/y_T}$,
with $\Lambda\simeq t$ the lattice cutoff. The
mean-field coefficient gives $E_T\simeq0.03$--$0.07$\,K, a coefficient of
the bond-level size $0.1$--$0.2$\,K ($0.2$--$0.5$\,K with $\Lambda=J'$), and one of order one
$0.5$--$1$\,K.

\emph{The field selection rule.}---The stabilizer of $(\text{DM},\bm B)$ contains the locked elements $O$ with
$O\bm B=\bm B$ and the products $\mathcal TO$ with $O\bm B=-\bm B$.
Classifying the 36 bilinears under it gives, for $\bm B\parallel[001]$,
only the flux singlet; for $\bm B\parallel[100]$ the N\'eel component
$N^y$; for $\bm B\parallel[110]$ the component $(N^x+N^y)/\sqrt2$ and, at
order $B_xB_y$, the valence-bond imbalance $V_x-V_y$. The order in the DM
couplings follows from $\hat\sigma_h$, which maps
$H(\bm D_\parallel,D'_\perp,\bm B)$ onto
$H(-\bm D_\parallel,D'_\perp,C_{2z}\bm B)$ and reverses $N^{x,y}$; the
locked group of the high-temperature structure, which contains
$\hat\sigma_h$ and therefore classifies the even orders, still allows the
tensor $(N^x,N^y)\propto(B_y,B_x)$. A Gaussian computation on the
$\pi$-flux state with the Wick-decoupled DM operator confirms the rule:
the N\'eel source is linear in $B$ and proportional to $\lambda D'_{\parallel,s}$, of second order in the in-plane couplings,
along $[010]$ for $\bm B\parallel[100]$ and along $[110]$ for
$\bm B\parallel[110]$.

Exact diagonalization of Eq.~\eqref{eq:H} with the DM vectors of the
$I\bar42m$ structure confirms the rule independently of the gauge theory,
on a 16-site cluster that keeps the full buckled group, a
20-site cluster that keeps $\hat S_4$, and an 8-site cluster at finite
temperature. On the 16-site cluster the linear response of $\bm N$ to $\bm B$ is the pure
off-diagonal tensor $(N^x,N^y)\propto(B_y,B_x)$, and every other entry is below
$10^{-15}$; it is even under $\bm D_\parallel\to-\bm D_\parallel$ to
$2\times10^{-14}$ and quadratic in the in-plane couplings; $D'_\perp$ alone gives
no response, and $\bm B\parallel c$ induces no N\'eel moment up to
$B=0.3J$. Its sign changes with $J'/J$ and with temperature. A $c$-axis field induces only the intra-dimer staggered moments of Ref.~\cite{Kodama2005}, and with a dimer-dependent $g$ tensor and no DM the N\'eel response equals $\tfrac12(g_1-g_2)\,\partial\langle\bm N\rangle/\partial h_s$.

No experiment has compared $[100]$ with $c$: Refs.~\cite{Cui2023,Cui2024two}
applied the field along $c$ or $8.6^\circ$ from it, in an uncharacterized
in-plane direction, and Ref.~\cite{Waki2007} compared $\bm B\parallel c$
with $\bm B\parallel[110]$ at $2.4$\,GPa at equal field. At $2.4$\,GPa
and $8$\,T along $[100]$, above $H_c$, Ref.~\cite{Wu2026} resolved no line
splitting down to $1.8$\,K, which bounds the $[100]$ response but does not
test its orientation dependence. The plaquette order
($T_P\simeq1.6$\,K at $2.4$\,GPa and $4$\,T~\cite{Cui2024two}) lowers the
layer symmetry, so the comparison belongs above $T_P$.
For a tilted field the conjugate staggered field is
$h_s\sim\tfrac12(g_1-g_2)\mu_BB\sin\theta$ plus a term of second order in
the in-plane DM couplings. It is relevant with eigenvalue $3-\Delta_N$ and
rounds a critical point on the scale
$T_h\sim\Lambda(h_s/\Lambda)^{1/(3-\Delta_N)}$.
For an illustrative
$|g_1-g_2|=10^{-2}$, the $8.6^\circ$ tilt of Ref.~\cite{Cui2023} at
$H_c\simeq5.7$\,T and a cutoff $\Lambda\simeq J'$--$J\simeq50$--$85$\,K give $T_h\sim0.1$\,K at
$\Delta_N=1.46$ and $\sim1$\,K at $\Delta_N\simeq0.63$.

\bibliography{arXiv_v1_PRL}

@article{ShastrySutherland,
  author = {Shastry, B. Sriram and Sutherland, Bill},
  title = {{Exact ground state of a quantum mechanical antiferromagnet}},
  journal = {Physica B+C},
  volume = {108},
  pages = {1069},
  year = {1981},
  doi = {10.1016/0378-4363(81)90838-X}
}

@article{Kageyama1999,
  author = {Kageyama, H. and Yoshimura, K. and Stern, R. and Mushnikov, N. V. and Onizuka, K. and Kato, M. and Kosuge, K. and Slichter, C. P. and Goto, T. and Ueda, Y.},
  title = {{Exact Dimer Ground State and Quantized Magnetization Plateaus in the Two-Dimensional Spin System SrCu$_2$(BO$_3$)$_2$}},
  journal = {Phys. Rev. Lett.},
  volume = {82},
  pages = {3168},
  year = {1999},
  doi = {10.1103/PhysRevLett.82.3168}
}

@article{MiyaharaUeda1999,
  author = {Miyahara, Shin and Ueda, Kazuo},
  title = {{Exact Dimer Ground State of the Two Dimensional Heisenberg Spin System SrCu$_2$(BO$_3$)$_2$}},
  journal = {Phys. Rev. Lett.},
  volume = {82},
  pages = {3701},
  year = {1999},
  doi = {10.1103/PhysRevLett.82.3701}
}

@article{MiyaharaUeda2003,
  author = {Miyahara, Shin and Ueda, Kazuo},
  title = {{Theory of the orthogonal dimer Heisenberg spin model for SrCu$_2$(BO$_3$)$_2$}},
  journal = {J. Phys.: Condens. Matter},
  volume = {15},
  pages = {R327},
  year = {2003},
  doi = {10.1088/0953-8984/15/9/201}
}

@article{LYSV2019,
  author = {Lee, Jong Yeon and You, Yi-Zhuang and Sachdev, Subir and Vishwanath, Ashvin},
  title = {{Signatures of a Deconfined Phase Transition on the Shastry-Sutherland Lattice: Applications to Quantum Critical SrCu$_2$(BO$_3$)$_2$}},
  journal = {Phys. Rev. X},
  volume = {9},
  pages = {041037},
  year = {2019},
  doi = {10.1103/PhysRevX.9.041037}
}

@article{Senthil2004a,
  author = {Senthil, T. and Vishwanath, Ashvin and Balents, Leon and Sachdev, Subir and Fisher, Matthew P. A.},
  title = {{Deconfined Quantum Critical Points}},
  journal = {Science},
  volume = {303},
  pages = {1490},
  year = {2004},
  doi = {10.1126/science.1091806}
}

@article{Senthil2004b,
  author = {Senthil, T. and Balents, Leon and Sachdev, Subir and Vishwanath, Ashvin and Fisher, Matthew P. A.},
  title = {{Quantum criticality beyond the Landau-Ginzburg-Wilson paradigm}},
  journal = {Phys. Rev. B},
  volume = {70},
  pages = {144407},
  year = {2004},
  doi = {10.1103/PhysRevB.70.144407}
}

@article{Zayed2017,
  author = {Zayed, M. E. and R{\"u}egg, Ch. and {Larrea J.}, J. and L{\"a}uchli, A. M. and Panagopoulos, C. and Saxena, S. S. and Ellerby, M. and McMorrow, D. F. and Str{\"a}ssle, Th. and Klotz, S. and Hamel, G. and Sadykov, R. A. and Pomjakushin, V. and Boehm, M. and Jim{\'e}nez-Ruiz, M. and Schneidewind, A. and Pomjakushina, E. and Stingaciu, M. and Conder, K. and R{\o}nnow, H. M.},
  title = {{4-spin plaquette singlet state in the Shastry--Sutherland compound SrCu$_2$(BO$_3$)$_2$}},
  journal = {Nat. Phys.},
  volume = {13},
  pages = {962},
  year = {2017},
  doi = {10.1038/nphys4190}
}

@article{Guo2020,
  author = {Guo, Jing and Sun, Guangyu and Zhao, Bowen and Wang, Ling and Hong, Wenshan and Sidorov, Vladimir A. and Ma, Nvsen and Wu, Qi and Li, Shiliang and Meng, Zi Yang and Sandvik, Anders W. and Sun, Liling},
  title = {{Quantum Phases of SrCu$_2$(BO$_3$)$_2$ from High-Pressure Thermodynamics}},
  journal = {Phys. Rev. Lett.},
  volume = {124},
  pages = {206602},
  year = {2020},
  doi = {10.1103/PhysRevLett.124.206602}
}

@article{LarreaJimenez2021,
  author = {{Larrea Jim{\'e}nez}, J. and Crone, S. P. G. and Fogh, E. and Zayed, M. E. and Lortz, R. and Pomjakushina, E. and Conder, K. and L{\"a}uchli, A. M. and Weber, L. and Wessel, S. and Honecker, A. and Normand, B. and R{\"u}egg, Ch. and Corboz, P. and R{\o}nnow, H. M. and Mila, F.},
  title = {{A quantum magnetic analogue to the critical point of water}},
  journal = {Nature},
  volume = {592},
  pages = {370},
  year = {2021},
  doi = {10.1038/s41586-021-03411-8}
}

@article{Cui2023,
  author = {Cui, Yi and Liu, Lu and Lin, Huihang and Wu, Kai-Hsin and Hong, Wenshan and Liu, Xuefei and Li, Cong and Hu, Ze and Xi, Ning and Li, Shiliang and Yu, Rong and Sandvik, Anders W. and Yu, Weiqiang},
  title = {{Proximate deconfined quantum critical point in SrCu$_2$(BO$_3$)$_2$}},
  journal = {Science},
  volume = {380},
  pages = {1179},
  year = {2023},
  doi = {10.1126/science.adc9487}
}

@article{Guo2025,
  author = {Guo, Jing and Wang, Pengyu and Huang, Cheng and Chen, Bin-Bin and Hong, Wenshan and Cai, Shu and Zhao, Jinyu and Han, Jinyu and Chen, Xintian and Zhou, Yazhou and Li, Shiliang and Wu, Qi and Meng, Zi Yang and Sun, Liling},
  title = {{Deconfined quantum critical point lost in pressurized SrCu$_2$(BO$_3$)$_2$}},
  journal = {Commun. Phys.},
  volume = {8},
  pages = {75},
  year = {2025},
  doi = {10.1038/s42005-025-01976-8}
}

@article{Guo2026,
  author = {Guo, Jing and Wang, Pengyu and Huang, Cheng and Zhou, Chengkang and Hong, Wenshan and Song, Menghan and Chen, Xintian and Wang, Ting-Tung and Cai, Shu and Zhao, Jinyu and Han, Jinyu and Zhou, Yazhou and Wu, Qi and Li, Shiliang and Meng, Zi Yang and Sun, Liling},
  title = {{$T$-linear specific heat in pressurized and magnetized Shastry-Sutherland Mott insulator SrCu$_2$(BO$_3$)$_2$}},
  journal = {Sci. Bull.},
  year = {2026},
  doi = {10.1016/j.scib.2026.09.060},
  note = {in press}
}

@misc{Song2026,
  author = {Song, Menghan and Zhou, Chengkang and Huang, Cheng and Meng, Zi Yang},
  title = {{Thermodynamics of Shastry-Sutherland Model under Magnetic Field}},
  year = {2026},
  archivePrefix = {arXiv},
  eprint = {2602.11589},
  doi = {10.48550/arXiv.2602.11589}
}

@article{Yang2022,
  author = {Yang, Jianwei and Sandvik, Anders W. and Wang, Ling},
  title = {{Quantum criticality and spin liquid phase in the Shastry-Sutherland model}},
  journal = {Phys. Rev. B},
  volume = {105},
  pages = {L060409},
  year = {2022},
  doi = {10.1103/PhysRevB.105.L060409}
}

@article{Viteritti2025,
  author = {Viteritti, Luciano Loris and Rende, Riccardo and Parola, Alberto and Goldt, Sebastian and Becca, Federico},
  title = {{Transformer wave function for two dimensional frustrated magnets: Emergence of a spin-liquid phase in the Shastry-Sutherland model}},
  journal = {Phys. Rev. B},
  volume = {111},
  pages = {134411},
  year = {2025},
  doi = {10.1103/PhysRevB.111.134411}
}

@article{Corboz2026,
  author = {Corboz, Philippe and Zhang, Yining and Ponsioen, Boris and Mila, Fr{\'e}d{\'e}ric},
  title = {{Quantum Spin Liquid Phase in the Shastry-Sutherland Model Revealed by High-Precision Infinite Projected Entangled-Pair States}},
  journal = {Phys. Rev. Lett.},
  volume = {136},
  pages = {186701},
  year = {2026},
  doi = {10.1103/67xm-5d8v}
}

@misc{Maity2025,
  author = {Maity, Atanu and Ferrari, Francesco and Capponi, Sylvain and Penc, Karlo and Potten, Janik and M{\"u}ller, Tobias and Feuerpfeil, Andreas and Thomale, Ronny and Lee, Jong Yeon and Samajdar, Rhine and Iqbal, Yasir},
  title = {{Evidence for a $\mathbb{Z}_{2}$ Dirac spin liquid in the generalized Shastry-Sutherland model}},
  year = {2025},
  archivePrefix = {arXiv},
  eprint = {2501.00096},
  doi = {10.48550/arXiv.2501.00096}
}

@article{Xi2023,
  author = {Xi, Ning and Chen, Hongyu and Xie, Z. Y. and Yu, Rong},
  title = {{Plaquette valence bond solid to antiferromagnet transition and deconfined quantum critical point of the Shastry-Sutherland model}},
  journal = {Phys. Rev. B},
  volume = {107},
  pages = {L220408},
  year = {2023},
  doi = {10.1103/PhysRevB.107.L220408}
}

@article{Qian-2025,
  title = {From the {Shastry-Sutherland} model to the ${J}_{1}\text{\ensuremath{-}}{J}_{2}$ {Heisenberg} model},
  author = {Qian, Xiangjian and Lv, Rongyi and Lee, Jong Yeon and Qin, Mingpu},
  journal = {Phys. Rev. B},
  volume = {111},
  issue = {24},
  pages = {L241113},
  numpages = {7},
  year = {2025},
  month = {Jun},
  publisher = {American Physical Society},
  doi = {10.1103/h6gc-t1wm},
  url = {https://link.aps.org/doi/10.1103/h6gc-t1wm}
}

@article{Wang2024,
  author = {Liu, Wen-Yuan and Zhang, Xiao-Tian and Wang, Zhe and Gong, Shou-Shu and Chen, Wei-Qiang and Gu, Zheng-Cheng},
  title = {{Quantum Criticality with Emergent Symmetry in the Extended Shastry-Sutherland Model}},
  journal = {Phys. Rev. Lett.},
  volume = {133},
  pages = {026502},
  year = {2024},
  doi = {10.1103/PhysRevLett.133.026502}
}

@article{ShackletonThomsonSachdev2021,
  author = {Shackleton, Leyna and Thomson, Alex and Sachdev, Subir},
  title = {{Deconfined criticality and a gapless $\mathbb{Z}_2$ spin liquid in the square-lattice antiferromagnet}},
  journal = {Phys. Rev. B},
  volume = {104},
  pages = {045110},
  year = {2021},
  doi = {10.1103/PhysRevB.104.045110}
}

@misc{Feuerpfeil2026,
  author = {Feuerpfeil, Andreas and Shackleton, Leyna and Maity, Atanu and Thomale, Ronny and Sachdev, Subir and Iqbal, Yasir},
  title = {{Unifying Dirac Spin Liquids on Square and Shastry-Sutherland Lattices via Fermionic Deconfined Criticality}},
  year = {2026},
  archivePrefix = {arXiv},
  eprint = {2601.19980},
  doi = {10.48550/arXiv.2601.19980},
  note = {accepted in SciPost Phys.; equation numbers refer to the accepted version}
}

@misc{Maity2026,
  author = {Maity, Atanu and Feuerpfeil, Andreas and Thomale, Ronny and Sachdev, Subir and Iqbal, Yasir},
  title = {{Unified gauge-theory description of quantum spin liquids on square-based frustrated lattices}},
  year = {2026},
  archivePrefix = {arXiv},
  eprint = {2603.15745},
  doi = {10.48550/arXiv.2603.15745}
}

@article{Cepas2001,
  author = {C{\'e}pas, O. and Kakurai, K. and Regnault, L. P. and Ziman, T. and Boucher, J. P. and Aso, N. and Nishi, M. and Kageyama, H. and Ueda, Y.},
  title = {{Dzyaloshinski-Moriya Interaction in the 2D Spin Gap System SrCu$_2$(BO$_3$)$_2$}},
  journal = {Phys. Rev. Lett.},
  volume = {87},
  pages = {167205},
  year = {2001},
  doi = {10.1103/PhysRevLett.87.167205}
}

@article{Kodama2005,
  author = {Kodama, K. and Miyahara, S. and Takigawa, M. and Horvati\'{c}, M. and Berthier, C. and Mila, F. and Kageyama, H. and Ueda, Y.},
  title = {{Field-induced effects of anisotropic magnetic interactions in SrCu$_2$(BO$_3$)$_2$}},
  journal = {J. Phys.: Condens. Matter},
  volume = {17},
  pages = {L61},
  year = {2005},
  doi = {10.1088/0953-8984/17/4/L02}
}

@article{Mazurenko2008,
  author = {Mazurenko, V. V. and Skornyakov, S. L. and Anisimov, V. I. and Mila, F.},
  title = {{First-principles investigation of symmetric and antisymmetric exchange interactions of SrCu$_2$(BO$_3$)$_2$}},
  journal = {Phys. Rev. B},
  volume = {78},
  pages = {195110},
  year = {2008},
  doi = {10.1103/PhysRevB.78.195110}
}

@article{Romhanyi2015,
  author = {Romh{\'a}nyi, Judit and Penc, Karlo and Ganesh, R.},
  title = {{Hall effect of triplons in a dimerized quantum magnet}},
  journal = {Nat. Commun.},
  volume = {6},
  pages = {6805},
  year = {2015},
  doi = {10.1038/ncomms7805}
}

@article{Fogh2024PRL,
  author = {Fogh, Ellen and Giriat, Ga{\'e}tan and Zayed, Mohamed E. and Piovano, Andrea and Boehm, Martin and Steffens, Paul and Safiulina, Irina and Hansen, Ursula B. and Klotz, Stefan and Soh, Jian-Rui and Pomjakushina, Ekaterina and Mila, Fr{\'e}d{\'e}ric and Normand, Bruce and R{\o}nnow, Henrik M.},
  title = {{Spin Waves and Three Dimensionality in the High-Pressure Antiferromagnetic Phase of SrCu$_2$(BO$_3$)$_2$}},
  journal = {Phys. Rev. Lett.},
  volume = {133},
  pages = {246702},
  year = {2024},
  doi = {10.1103/PhysRevLett.133.246702}
}

@article{Cederholm2026,
  author = {Cederholm, Jeppe Jon and Piovano, Andrea and Ivanov, Alexandre and Klotz, Stefan and Hiess, Arno and Hansen, Ursula B. and Pomjakushina, Ekaterina and Zayed, Mohamed E. and Fogh, Ellen and R{\o}nnow, Henrik M.},
  title = {{Spin excitations near the pressure-induced antiferromagnetic transition in SrCu$_2$(BO$_3$)$_2$}},
  journal = {J. Appl. Crystallogr.},
  volume = {59},
  pages = {804},
  year = {2026},
  doi = {10.1107/S1600576726003705}
}

@misc{Cui2024two,
  author = {Cui, Yi and Du, Kefan and Wu, Zhanlong and Li, Shuo and Yang, Pengtao and Chen, Ying and Xu, Xiaoyu and Chen, Hongyu and Li, Chengchen and Liu, Juanjuan and Wang, Bosen and Hong, Wenshan and Li, Shiliang and Xie, Zhiyuan and Cheng, Jinguang and Normand, Bruce and Yu, Rong and Yu, Weiqiang},
  title = {{Two Plaquette-Singlet Phases and Emergent SO(5) Deconfined Quantum Criticality in SrCu$_2$(BO$_3$)$_2$}},
  year = {2024},
  archivePrefix = {arXiv},
  eprint = {2411.00302},
  doi = {10.48550/arXiv.2411.00302}
}

@article{Hermele2008,
  author = {Hermele, Michael and Ran, Ying and Lee, Patrick A. and Wen, Xiao-Gang},
  title = {{Properties of an algebraic spin liquid on the kagome lattice}},
  journal = {Phys. Rev. B},
  volume = {77},
  pages = {224413},
  year = {2008},
  doi = {10.1103/PhysRevB.77.224413}
}

@article{Cepas2008,
  author = {C{\'e}pas, O. and Fong, C. M. and Leung, P. W. and Lhuillier, C.},
  title = {{Quantum phase transition induced by Dzyaloshinskii-Moriya interactions in the kagome antiferromagnet}},
  journal = {Phys. Rev. B},
  volume = {78},
  pages = {140405(R)},
  year = {2008},
  doi = {10.1103/PhysRevB.78.140405}
}

@article{Kagome2407,
  author = {He, Li-Wei and Li, Jian-Xin},
  title = {{Spinon quantum spin Hall state in the kagome antiferromagnet with a Dzyaloshinskii-Moriya interaction}},
  journal = {Phys. Rev. B},
  volume = {110},
  pages = {035131},
  year = {2024},
  doi = {10.1103/PhysRevB.110.035131}
}

@article{Ferrari-2023,
	title = {{Static and dynamical signatures of Dzyaloshinskii-Moriya interactions in the Heisenberg model on the kagome lattice}},
	pages = {139},
	author = {Ferrari, Francesco and Niu, Sen and Hasik, Juraj and Iqbal, Yasir and Poilblanc, Didier and Becca, Federico},
	journal = {SciPost Phys.},
	volume = {14},
	year = {2023},
	publisher = {SciPost},
	doi = {10.21468/SciPostPhys.14.6.139},
	url = {https://scipost.org/10.21468/SciPostPhys.14.6.139}
}

@article{Huh2010,
  author = {Huh, Yejin and Fritz, Lars and Sachdev, Subir},
  title = {{Quantum criticality of the kagome antiferromagnet with Dzyaloshinskii-Moriya interactions}},
  journal = {Phys. Rev. B},
  volume = {81},
  pages = {144432},
  year = {2010},
  doi = {10.1103/PhysRevB.81.144432}
}

@article{Sparta2001,
  author = {Sparta, K. and Redhammer, G. J. and Roussel, P. and Heger, G. and Roth, G. and Lemmens, P. and Ionescu, A. and Grove, M. and G{\"u}ntherodt, G. and H{\"u}ning, F. and Lueken, H. and Kageyama, H. and Onizuka, K. and Ueda, Y.},
  title = {{Structural phase transition in the 2D spin dimer compound SrCu$_2$(BO$_3$)$_2$}},
  journal = {Eur. Phys. J. B},
  volume = {19},
  pages = {507},
  year = {2001},
  doi = {10.1007/s100510170296}
}

@article{Loa2005,
  author = {Loa, I. and Zhang, F. X. and Syassen, K. and Lemmens, P. and Crichton, W. and Kageyama, H. and Ueda, Y.},
  title = {{Crystal structure and lattice dynamics of SrCu$_2$(BO$_3$)$_2$ at high pressures}},
  journal = {Physica B},
  volume = {359--361},
  pages = {980},
  year = {2005},
  doi = {10.1016/j.physb.2005.01.284}
}

@article{Zayed2014,
  author = {Zayed, M. E. and R\"uegg, Ch. and Pomjakushina, E. and Stingaciu, M. and Conder, K. and Hanfland, M. and Merlini, M. and R{\o}nnow, H. M.},
  title = {{Temperature dependence of the pressure induced monoclinic distortion in the spin $S=1/2$ Shastry--Sutherland compound SrCu$_2$(BO$_3$)$_2$}},
  journal = {Solid State Commun.},
  volume = {186},
  pages = {13},
  year = {2014},
  doi = {10.1016/j.ssc.2014.01.008}
}

@article{Cheng2007,
  author = {Cheng, Y. F. and C{\'e}pas, O. and Leung, P. W. and Ziman, T.},
  title = {{Magnon dispersion and anisotropies in SrCu$_2$(BO$_3$)$_2$}},
  journal = {Phys. Rev. B},
  volume = {75},
  pages = {144422},
  year = {2007},
  doi = {10.1103/PhysRevB.75.144422}
}

@unpublished{companion,
  author = {Iqbal, Yasir},
  title  = {{Dzyaloshinskii--Moriya interactions in the SU(2) gauge theory of the Shastry--Sutherland Dirac spin liquids}},
  note   = {companion paper (Physical Review B, submitted together with this Letter)},
  year   = {2026}
}

@article{ThomsonSachdev,
  author = {Thomson, Alex and Sachdev, Subir},
  title = {{Fermionic Spinon Theory of Square Lattice Spin Liquids near the N\'{e}el State}},
  journal = {Phys. Rev. X},
  volume = {8},
  pages = {011012},
  year = {2018},
  doi = {10.1103/PhysRevX.8.011012}
}

@article{WangNahum2017,
  author = {Wang, Chong and Nahum, Adam and Metlitski, Max A. and Xu, Cenke and Senthil, T.},
  title = {{Deconfined Quantum Critical Points: Symmetries and Dualities}},
  journal = {Phys. Rev. X},
  volume = {7},
  pages = {031051},
  year = {2017},
  doi = {10.1103/PhysRevX.7.031051}
}

@misc{SM,
  note = {See the Supplemental Material at the end of this article}
}

@article{Sun2009,
  author = {Sun, Kai and Yao, Hong and Fradkin, Eduardo and Kivelson, Steven A.},
  title = {{Topological Insulators and Nematic Phases from Spontaneous Symmetry Breaking in 2D Fermi Systems with a Quadratic Band Crossing}},
  journal = {Phys. Rev. Lett.},
  volume = {103},
  pages = {046811},
  year = {2009},
  doi = {10.1103/PhysRevLett.103.046811}
}

@article{VafekYang2010,
  author = {Vafek, Oskar and Yang, Kun},
  title = {{Many-body instability of Coulomb interacting bilayer graphene: Renormalization group approach}},
  journal = {Phys. Rev. B},
  volume = {81},
  pages = {041401(R)},
  year = {2010},
  doi = {10.1103/PhysRevB.81.041401}
}

@article{Minwalla1998,
  author = {Minwalla, Shiraz},
  title = {{Restrictions imposed by superconformal invariance on quantum field theories}},
  journal = {Adv. Theor. Math. Phys.},
  volume = {2},
  pages = {783},
  year = {1998},
  doi = {10.4310/ATMP.1998.v2.n4.a4}
}

@article{Evans1967,
  author = {Evans, N. T.},
  title = {{Discrete Series for the Universal Covering Group of the 3+2 de Sitter Group}},
  journal = {J. Math. Phys.},
  volume = {8},
  pages = {170},
  year = {1967},
  doi = {10.1063/1.1705183}
}

@article{Zhou2024fuzzy,
  author = {Zhou, Zheng and Hu, Liangdong and Zhu, W. and He, Yin-Chen},
  title = {{SO(5) Deconfined Phase Transition under the Fuzzy-Sphere Microscope: Approximate Conformal Symmetry, Pseudo-Criticality, and Operator Spectrum}},
  journal = {Phys. Rev. X},
  volume = {14},
  pages = {021044},
  year = {2024},
  doi = {10.1103/PhysRevX.14.021044}
}

@article{OshikawaAffleck1997,
  author = {Oshikawa, Masaki and Affleck, Ian},
  title = {{Field-Induced Gap in $S=1/2$ Antiferromagnetic Chains}},
  journal = {Phys. Rev. Lett.},
  volume = {79},
  pages = {2883},
  year = {1997},
  doi = {10.1103/PhysRevLett.79.2883}
}

@article{Miyahara2004,
  author = {Miyahara, S. and Mila, F. and Kodama, K. and Takigawa, M. and Horvati{\'c}, M. and Berthier, C. and Kageyama, H. and Ueda, Y.},
  title = {{The effects of intra-dimer Dzyaloshinsky-Moriya interaction on the properties of SrCu$_2$(BO$_3$)$_2$ in an external magnetic field}},
  journal = {J. Phys.: Condens. Matter},
  volume = {16},
  pages = {S911},
  year = {2004},
  doi = {10.1088/0953-8984/16/11/048}
}

@article{Miyahara2007,
  author  = {Miyahara, S. and Fouet, J.-B. and Manmana, S. R. and Noack, R. M. and Mayaffre, H. and Sheikin, I. and Berthier, C. and Mila, F.},
  title   = {{Uniform and staggered magnetizations induced by Dzyaloshinskii-Moriya interactions in isolated and coupled spin-1/2 dimers in a magnetic field}},
  journal = {Phys. Rev. B},
  volume  = {75},
  pages   = {184402},
  year    = {2007},
  doi     = {10.1103/PhysRevB.75.184402}
}

@article{Nahum2015,
  author = {Nahum, Adam and Chalker, J. T. and Serna, P. and Ortu{\~n}o, M. and Somoza, A. M.},
  title = {{Deconfined Quantum Criticality, Scaling Violations, and Classical Loop Models}},
  journal = {Phys. Rev. X},
  volume = {5},
  pages = {041048},
  year = {2015},
  doi = {10.1103/PhysRevX.5.041048}
}

@misc{MazurenkoPC,
  author = {Mazurenko, V. V.},
  note = {private communication (2026)}
}

@article{ZhouHe2023,
  author = {Zhou, Zheng and He, Yin-Chen},
  title = {{Slightly broken higher-spin current in bosonic and fermionic QED in the large-$N$ limit}},
  journal = {SciPost Phys.},
  volume = {15},
  pages = {072},
  year = {2023},
  doi = {10.21468/SciPostPhys.15.2.072}
}

@article{Gracey1994,
  author = {Gracey, John A.},
  title = {{Anomalous dimension of non-singlet Wilson operators at $O(1/N_f)$ in deep inelastic scattering}},
  journal = {Phys. Lett. B},
  volume = {322},
  pages = {141},
  year = {1994},
  doi = {10.1016/0370-2693(94)90502-9}
}

@article{RantnerWen2002,
  author = {Rantner, Walter and Wen, Xiao-Gang},
  title = {{Spin correlations in the algebraic spin liquid: Implications for high-$T_c$ superconductors}},
  journal = {Phys. Rev. B},
  volume = {66},
  pages = {144501},
  year = {2002},
  doi = {10.1103/PhysRevB.66.144501}
}

@article{Hermele2005,
  author = {Hermele, Michael and Senthil, T. and Fisher, Matthew P. A.},
  title = {{Algebraic spin liquid as the mother of many competing orders}},
  journal = {Phys. Rev. B},
  volume = {72},
  pages = {104404},
  year = {2005},
  doi = {10.1103/PhysRevB.72.104404}
}

@article{Shekhtman1992,
  author = {Shekhtman, L. and Entin-Wohlman, O. and Aharony, Amnon},
  title = {{Moriya's Anisotropic Superexchange Interaction, Frustration, and Dzyaloshinsky's Weak Ferromagnetism}},
  journal = {Phys. Rev. Lett.},
  volume = {69},
  pages = {836},
  year = {1992},
  doi = {10.1103/PhysRevLett.69.836}
}

@article{Waki2007,
  author = {Waki, Takeshi and Arai, Koichi and Takigawa, Masashi and Saiga, Yuta and Uwatoko, Yoshiya and Kageyama, Hiroshi and Ueda, Yutaka},
  title = {{A Novel Ordered Phase in SrCu$_2$(BO$_3$)$_2$ under High Pressure}},
  journal = {J. Phys. Soc. Jpn.},
  volume = {76},
  pages = {073710},
  year = {2007},
  doi = {10.1143/JPSJ.76.073710}
}

@article{Wu2026,
  author = {Wu, Zhanlong and Du, Kefan and Li, Shuo and Shi, Tong and Chen, Ying and Dong, Qingxin and Zhou, Rui and Yu, Rong and Liu, Juanjuan and Wang, Bosen and Cheng, Jinguang and Yu, Weiqiang and Cui, Yi},
  title = {{NMR evidence of pressure-induced structural transition and enhanced spin fluctuations up to 14 GPa in SrCu$_2$(BO$_3$)$_2$}},
  journal = {Phys. Rev. B},
  volume = {114},
  pages = {L231101},
  year = {2026},
  doi = {10.1103/6bxt-c4zm}
}

\clearpage
\onecolumngrid
\setcounter{table}{0}
\setcounter{figure}{0}
\setcounter{equation}{0}
\renewcommand{\thetable}{S\arabic{table}}
\renewcommand{\thefigure}{S\arabic{figure}}
\renewcommand{\theequation}{S\arabic{equation}}
\begin{center}
{\large\bfseries Supplemental Material}
\end{center}
This Supplemental Material lists the transformation tables on which the
no-mass theorem and the operator selection of the Letter rest, and the
bond-resolved Dzyaloshinskii--Moriya (DM) textures. Derivations and
all further details, including the one-loop and $O(1/N)$ results quoted in the Letter, are given in the companion paper~\cite{companion}.
Rows are labeled by the matrix $M\sigma^b$ of the gauge-singlet bilinear
$\Tr[\Xb M\X\sigma^b]$ (notation of the Letter); the entries give the image under each generator
($\pm$ alone means $\pm$ the same row). The tables use spin labels
related to physical axes by $x\to-x$; this relabeling does not change
any sign on the diagonal.

\begin{table}[ht]
\caption{All 36 gauge-singlet bilinears under the unlocked generators of
the Shastry--Sutherland lattice (no spin--orbit coupling). First block:
the six masses (SO(5) vector and flux singlet). Second block: the 30
conserved SO(5)-adjoint current components.}
\begin{ruledtabular}
\begin{tabular}{lccccc}
$\mathrm{Tr}[\bar{\mathcal X}\,\cdot\,\mathcal X]$ & $G_x$ & $\sigma_{xy}$ & $\sigma_{x\bar y}$ & $C_4$ & $\mathcal T$ \\
\colrule
$\one$ & $-$ & $-$ & $-$ & $+$ & $-$ \\
$\mu^x$ & $-$ & $-\mu^z$ & $-\mu^z$ & $+\mu^z$ & $+$ \\
$\mu^y\sigma^x$ & $-$ & $+$ & $+$ & $-$ & $-$ \\
$\mu^y\sigma^y$ & $-$ & $+$ & $+$ & $-$ & $-$ \\
$\mu^y\sigma^z$ & $-$ & $+$ & $+$ & $-$ & $-$ \\
$\mu^z$ & $-$ & $-\mu^x$ & $-\mu^x$ & $+\mu^x$ & $+$ \\
\colrule
$\gamma^0\sigma^x$ & $+$ & $+$ & $+$ & $+$ & $-$ \\
$\gamma^x\sigma^x$ & $+$ & $+\gamma^y\sigma^x$ & $-\gamma^y\sigma^x$ & $-\gamma^y\sigma^x$ & $+$ \\
$\gamma^y\sigma^x$ & $-$ & $+\gamma^x\sigma^x$ & $-\gamma^x\sigma^x$ & $+\gamma^x\sigma^x$ & $+$ \\
$\gamma^0\sigma^y$ & $+$ & $+$ & $+$ & $+$ & $-$ \\
$\gamma^x\sigma^y$ & $+$ & $+\gamma^y\sigma^y$ & $-\gamma^y\sigma^y$ & $-\gamma^y\sigma^y$ & $+$ \\
$\gamma^y\sigma^y$ & $-$ & $+\gamma^x\sigma^y$ & $-\gamma^x\sigma^y$ & $+\gamma^x\sigma^y$ & $+$ \\
$\gamma^0\sigma^z$ & $+$ & $+$ & $+$ & $+$ & $-$ \\
$\gamma^x\sigma^z$ & $+$ & $+\gamma^y\sigma^z$ & $-\gamma^y\sigma^z$ & $-\gamma^y\sigma^z$ & $+$ \\
$\gamma^y\sigma^z$ & $-$ & $+\gamma^x\sigma^z$ & $-\gamma^x\sigma^z$ & $+\gamma^x\sigma^z$ & $+$ \\
$\gamma^0\mu^x\sigma^x$ & $+$ & $+\gamma^0\mu^z\sigma^x$ & $+\gamma^0\mu^z\sigma^x$ & $+\gamma^0\mu^z\sigma^x$ & $+$ \\
$\gamma^x\mu^x\sigma^x$ & $+$ & $+\gamma^y\mu^z\sigma^x$ & $-\gamma^y\mu^z\sigma^x$ & $-\gamma^y\mu^z\sigma^x$ & $-$ \\
$\gamma^y\mu^x\sigma^x$ & $-$ & $+\gamma^x\mu^z\sigma^x$ & $-\gamma^x\mu^z\sigma^x$ & $+\gamma^x\mu^z\sigma^x$ & $-$ \\
$\gamma^0\mu^x\sigma^y$ & $+$ & $+\gamma^0\mu^z\sigma^y$ & $+\gamma^0\mu^z\sigma^y$ & $+\gamma^0\mu^z\sigma^y$ & $+$ \\
$\gamma^x\mu^x\sigma^y$ & $+$ & $+\gamma^y\mu^z\sigma^y$ & $-\gamma^y\mu^z\sigma^y$ & $-\gamma^y\mu^z\sigma^y$ & $-$ \\
$\gamma^y\mu^x\sigma^y$ & $-$ & $+\gamma^x\mu^z\sigma^y$ & $-\gamma^x\mu^z\sigma^y$ & $+\gamma^x\mu^z\sigma^y$ & $-$ \\
$\gamma^0\mu^x\sigma^z$ & $+$ & $+\gamma^0\mu^z\sigma^z$ & $+\gamma^0\mu^z\sigma^z$ & $+\gamma^0\mu^z\sigma^z$ & $+$ \\
$\gamma^x\mu^x\sigma^z$ & $+$ & $+\gamma^y\mu^z\sigma^z$ & $-\gamma^y\mu^z\sigma^z$ & $-\gamma^y\mu^z\sigma^z$ & $-$ \\
$\gamma^y\mu^x\sigma^z$ & $-$ & $+\gamma^x\mu^z\sigma^z$ & $-\gamma^x\mu^z\sigma^z$ & $+\gamma^x\mu^z\sigma^z$ & $-$ \\
$\gamma^0\mu^y$ & $+$ & $-$ & $-$ & $-$ & $-$ \\
$\gamma^x\mu^y$ & $+$ & $-\gamma^y\mu^y$ & $+\gamma^y\mu^y$ & $+\gamma^y\mu^y$ & $+$ \\
$\gamma^y\mu^y$ & $-$ & $-\gamma^x\mu^y$ & $+\gamma^x\mu^y$ & $-\gamma^x\mu^y$ & $+$ \\
$\gamma^0\mu^z\sigma^x$ & $+$ & $+\gamma^0\mu^x\sigma^x$ & $+\gamma^0\mu^x\sigma^x$ & $+\gamma^0\mu^x\sigma^x$ & $+$ \\
$\gamma^x\mu^z\sigma^x$ & $+$ & $+\gamma^y\mu^x\sigma^x$ & $-\gamma^y\mu^x\sigma^x$ & $-\gamma^y\mu^x\sigma^x$ & $-$ \\
$\gamma^y\mu^z\sigma^x$ & $-$ & $+\gamma^x\mu^x\sigma^x$ & $-\gamma^x\mu^x\sigma^x$ & $+\gamma^x\mu^x\sigma^x$ & $-$ \\
$\gamma^0\mu^z\sigma^y$ & $+$ & $+\gamma^0\mu^x\sigma^y$ & $+\gamma^0\mu^x\sigma^y$ & $+\gamma^0\mu^x\sigma^y$ & $+$ \\
$\gamma^x\mu^z\sigma^y$ & $+$ & $+\gamma^y\mu^x\sigma^y$ & $-\gamma^y\mu^x\sigma^y$ & $-\gamma^y\mu^x\sigma^y$ & $-$ \\
$\gamma^y\mu^z\sigma^y$ & $-$ & $+\gamma^x\mu^x\sigma^y$ & $-\gamma^x\mu^x\sigma^y$ & $+\gamma^x\mu^x\sigma^y$ & $-$ \\
$\gamma^0\mu^z\sigma^z$ & $+$ & $+\gamma^0\mu^x\sigma^z$ & $+\gamma^0\mu^x\sigma^z$ & $+\gamma^0\mu^x\sigma^z$ & $+$ \\
$\gamma^x\mu^z\sigma^z$ & $+$ & $+\gamma^y\mu^x\sigma^z$ & $-\gamma^y\mu^x\sigma^z$ & $-\gamma^y\mu^x\sigma^z$ & $-$ \\
$\gamma^y\mu^z\sigma^z$ & $-$ & $+\gamma^x\mu^x\sigma^z$ & $-\gamma^x\mu^x\sigma^z$ & $+\gamma^x\mu^x\sigma^z$ & $-$ \\
\end{tabular}
\end{ruledtabular}
\end{table}

\begin{table}[ht]
\caption{The same under the spin--orbit-locked generators of the buckled
$I\bar42m$ structure [Eq.~(2) of the Letter, and $\hat\sigma_{x\bar y}=\sigma_{x\bar y}\otimes C_2^{[110]}$]. No mass row is invariant:
the N\'eel and flux masses are odd under $\mathcal T$, the valence-bond-solid (VBS) masses
$\mu^x,\mu^z$ under $\hat G_x$. The current rows $\gamma^x\sigma^x$ and
$\gamma^y\sigma^y$ map into each other with unit signs, so their sum,
$\mathcal O_{\rm DM}$, is the unique invariant. In physical axes
($\sigma^x_{\rm label}=-S^x$) it reads $J^x_x-J^y_y$ up to sign. Without $\mathcal T$
two further invariants exist, both $\mathcal T$-odd currents; the locked
space group alone, e.g.\ $\{\hat G_x,\hat\sigma_{xy}\}$, leaves no mass
invariant.}
\begin{ruledtabular}
\begin{tabular}{lccccc}
$\mathrm{Tr}[\bar{\mathcal X}\,\cdot\,\mathcal X]$ & $\hat G_x$ & $\hat\sigma_{xy}$ & $\hat\sigma_{x\bar y}$ & $\hat S_4$ & $\mathcal T$ \\
\colrule
$\one$ & $-$ & $-$ & $-$ & $+$ & $-$ \\
$\mu^x$ & $-$ & $-\mu^z$ & $-\mu^z$ & $+\mu^z$ & $+$ \\
$\mu^y\sigma^x$ & $-$ & $+\mu^y\sigma^y$ & $-\mu^y\sigma^y$ & $+\mu^y\sigma^y$ & $-$ \\
$\mu^y\sigma^y$ & $+$ & $+\mu^y\sigma^x$ & $-\mu^y\sigma^x$ & $-\mu^y\sigma^x$ & $-$ \\
$\mu^y\sigma^z$ & $+$ & $-$ & $-$ & $-$ & $-$ \\
$\mu^z$ & $-$ & $-\mu^x$ & $-\mu^x$ & $+\mu^x$ & $+$ \\
\colrule
$\gamma^0\sigma^x$ & $+$ & $+\gamma^0\sigma^y$ & $-\gamma^0\sigma^y$ & $-\gamma^0\sigma^y$ & $-$ \\
$\gamma^x\sigma^x$ & $+$ & $+\gamma^y\sigma^y$ & $+\gamma^y\sigma^y$ & $+\gamma^y\sigma^y$ & $+$ \\
$\gamma^y\sigma^x$ & $-$ & $+\gamma^x\sigma^y$ & $+\gamma^x\sigma^y$ & $-\gamma^x\sigma^y$ & $+$ \\
$\gamma^0\sigma^y$ & $-$ & $+\gamma^0\sigma^x$ & $-\gamma^0\sigma^x$ & $+\gamma^0\sigma^x$ & $-$ \\
$\gamma^x\sigma^y$ & $-$ & $+\gamma^y\sigma^x$ & $+\gamma^y\sigma^x$ & $-\gamma^y\sigma^x$ & $+$ \\
$\gamma^y\sigma^y$ & $+$ & $+\gamma^x\sigma^x$ & $+\gamma^x\sigma^x$ & $+\gamma^x\sigma^x$ & $+$ \\
$\gamma^0\sigma^z$ & $-$ & $-$ & $-$ & $+$ & $-$ \\
$\gamma^x\sigma^z$ & $-$ & $-\gamma^y\sigma^z$ & $+\gamma^y\sigma^z$ & $-\gamma^y\sigma^z$ & $+$ \\
$\gamma^y\sigma^z$ & $+$ & $-\gamma^x\sigma^z$ & $+\gamma^x\sigma^z$ & $+\gamma^x\sigma^z$ & $+$ \\
$\gamma^0\mu^x\sigma^x$ & $+$ & $+\gamma^0\mu^z\sigma^y$ & $-\gamma^0\mu^z\sigma^y$ & $-\gamma^0\mu^z\sigma^y$ & $+$ \\
$\gamma^x\mu^x\sigma^x$ & $+$ & $+\gamma^y\mu^z\sigma^y$ & $+\gamma^y\mu^z\sigma^y$ & $+\gamma^y\mu^z\sigma^y$ & $-$ \\
$\gamma^y\mu^x\sigma^x$ & $-$ & $+\gamma^x\mu^z\sigma^y$ & $+\gamma^x\mu^z\sigma^y$ & $-\gamma^x\mu^z\sigma^y$ & $-$ \\
$\gamma^0\mu^x\sigma^y$ & $-$ & $+\gamma^0\mu^z\sigma^x$ & $-\gamma^0\mu^z\sigma^x$ & $+\gamma^0\mu^z\sigma^x$ & $+$ \\
$\gamma^x\mu^x\sigma^y$ & $-$ & $+\gamma^y\mu^z\sigma^x$ & $+\gamma^y\mu^z\sigma^x$ & $-\gamma^y\mu^z\sigma^x$ & $-$ \\
$\gamma^y\mu^x\sigma^y$ & $+$ & $+\gamma^x\mu^z\sigma^x$ & $+\gamma^x\mu^z\sigma^x$ & $+\gamma^x\mu^z\sigma^x$ & $-$ \\
$\gamma^0\mu^x\sigma^z$ & $-$ & $-\gamma^0\mu^z\sigma^z$ & $-\gamma^0\mu^z\sigma^z$ & $+\gamma^0\mu^z\sigma^z$ & $+$ \\
$\gamma^x\mu^x\sigma^z$ & $-$ & $-\gamma^y\mu^z\sigma^z$ & $+\gamma^y\mu^z\sigma^z$ & $-\gamma^y\mu^z\sigma^z$ & $-$ \\
$\gamma^y\mu^x\sigma^z$ & $+$ & $-\gamma^x\mu^z\sigma^z$ & $+\gamma^x\mu^z\sigma^z$ & $+\gamma^x\mu^z\sigma^z$ & $-$ \\
$\gamma^0\mu^y$ & $+$ & $-$ & $-$ & $-$ & $-$ \\
$\gamma^x\mu^y$ & $+$ & $-\gamma^y\mu^y$ & $+\gamma^y\mu^y$ & $+\gamma^y\mu^y$ & $+$ \\
$\gamma^y\mu^y$ & $-$ & $-\gamma^x\mu^y$ & $+\gamma^x\mu^y$ & $-\gamma^x\mu^y$ & $+$ \\
$\gamma^0\mu^z\sigma^x$ & $+$ & $+\gamma^0\mu^x\sigma^y$ & $-\gamma^0\mu^x\sigma^y$ & $-\gamma^0\mu^x\sigma^y$ & $+$ \\
$\gamma^x\mu^z\sigma^x$ & $+$ & $+\gamma^y\mu^x\sigma^y$ & $+\gamma^y\mu^x\sigma^y$ & $+\gamma^y\mu^x\sigma^y$ & $-$ \\
$\gamma^y\mu^z\sigma^x$ & $-$ & $+\gamma^x\mu^x\sigma^y$ & $+\gamma^x\mu^x\sigma^y$ & $-\gamma^x\mu^x\sigma^y$ & $-$ \\
$\gamma^0\mu^z\sigma^y$ & $-$ & $+\gamma^0\mu^x\sigma^x$ & $-\gamma^0\mu^x\sigma^x$ & $+\gamma^0\mu^x\sigma^x$ & $+$ \\
$\gamma^x\mu^z\sigma^y$ & $-$ & $+\gamma^y\mu^x\sigma^x$ & $+\gamma^y\mu^x\sigma^x$ & $-\gamma^y\mu^x\sigma^x$ & $-$ \\
$\gamma^y\mu^z\sigma^y$ & $+$ & $+\gamma^x\mu^x\sigma^x$ & $+\gamma^x\mu^x\sigma^x$ & $+\gamma^x\mu^x\sigma^x$ & $-$ \\
$\gamma^0\mu^z\sigma^z$ & $-$ & $-\gamma^0\mu^x\sigma^z$ & $-\gamma^0\mu^x\sigma^z$ & $+\gamma^0\mu^x\sigma^z$ & $+$ \\
$\gamma^x\mu^z\sigma^z$ & $-$ & $-\gamma^y\mu^x\sigma^z$ & $+\gamma^y\mu^x\sigma^z$ & $-\gamma^y\mu^x\sigma^z$ & $-$ \\
$\gamma^y\mu^z\sigma^z$ & $+$ & $-\gamma^x\mu^x\sigma^z$ & $+\gamma^x\mu^x\sigma^z$ & $+\gamma^x\mu^x\sigma^z$ & $-$ \\
\end{tabular}
\end{ruledtabular}
\end{table}

\begin{table}[ht]
\caption{The two generators that the high-temperature structure adds, the
mirror $\hat\sigma_h=\one\otimes C_{2z}$ and the pure fourfold rotation
$\hat C_4=C_4\otimes R_z(+\pi/2)$. With the spin U(1)$_z$ of $D'_\perp$
alone, no bilinear is invariant.}
\begin{ruledtabular}
\begin{tabular}{lcc}
$\mathrm{Tr}[\bar{\mathcal X}\,\cdot\,\mathcal X]$ & $\hat\sigma_{h}$ & $\hat C_4$ \\
\colrule
$\one$ & $+$ & $+$ \\
$\mu^x$ & $+$ & $+\mu^z$ \\
$\mu^y\sigma^x$ & $-$ & $-\mu^y\sigma^y$ \\
$\mu^y\sigma^y$ & $-$ & $+\mu^y\sigma^x$ \\
$\mu^y\sigma^z$ & $+$ & $-$ \\
$\mu^z$ & $+$ & $+\mu^x$ \\
\colrule
$\gamma^0\sigma^x$ & $-$ & $+\gamma^0\sigma^y$ \\
$\gamma^x\sigma^x$ & $-$ & $-\gamma^y\sigma^y$ \\
$\gamma^y\sigma^x$ & $-$ & $+\gamma^x\sigma^y$ \\
$\gamma^0\sigma^y$ & $-$ & $-\gamma^0\sigma^x$ \\
$\gamma^x\sigma^y$ & $-$ & $+\gamma^y\sigma^x$ \\
$\gamma^y\sigma^y$ & $-$ & $-\gamma^x\sigma^x$ \\
$\gamma^0\sigma^z$ & $+$ & $+$ \\
$\gamma^x\sigma^z$ & $+$ & $-\gamma^y\sigma^z$ \\
$\gamma^y\sigma^z$ & $+$ & $+\gamma^x\sigma^z$ \\
$\gamma^0\mu^x\sigma^x$ & $-$ & $+\gamma^0\mu^z\sigma^y$ \\
$\gamma^x\mu^x\sigma^x$ & $-$ & $-\gamma^y\mu^z\sigma^y$ \\
$\gamma^y\mu^x\sigma^x$ & $-$ & $+\gamma^x\mu^z\sigma^y$ \\
$\gamma^0\mu^x\sigma^y$ & $-$ & $-\gamma^0\mu^z\sigma^x$ \\
$\gamma^x\mu^x\sigma^y$ & $-$ & $+\gamma^y\mu^z\sigma^x$ \\
$\gamma^y\mu^x\sigma^y$ & $-$ & $-\gamma^x\mu^z\sigma^x$ \\
$\gamma^0\mu^x\sigma^z$ & $+$ & $+\gamma^0\mu^z\sigma^z$ \\
$\gamma^x\mu^x\sigma^z$ & $+$ & $-\gamma^y\mu^z\sigma^z$ \\
$\gamma^y\mu^x\sigma^z$ & $+$ & $+\gamma^x\mu^z\sigma^z$ \\
$\gamma^0\mu^y$ & $+$ & $-$ \\
$\gamma^x\mu^y$ & $+$ & $+\gamma^y\mu^y$ \\
$\gamma^y\mu^y$ & $+$ & $-\gamma^x\mu^y$ \\
$\gamma^0\mu^z\sigma^x$ & $-$ & $+\gamma^0\mu^x\sigma^y$ \\
$\gamma^x\mu^z\sigma^x$ & $-$ & $-\gamma^y\mu^x\sigma^y$ \\
$\gamma^y\mu^z\sigma^x$ & $-$ & $+\gamma^x\mu^x\sigma^y$ \\
$\gamma^0\mu^z\sigma^y$ & $-$ & $-\gamma^0\mu^x\sigma^x$ \\
$\gamma^x\mu^z\sigma^y$ & $-$ & $+\gamma^y\mu^x\sigma^x$ \\
$\gamma^y\mu^z\sigma^y$ & $-$ & $-\gamma^x\mu^x\sigma^x$ \\
$\gamma^0\mu^z\sigma^z$ & $+$ & $+\gamma^0\mu^x\sigma^z$ \\
$\gamma^x\mu^z\sigma^z$ & $+$ & $-\gamma^y\mu^x\sigma^z$ \\
$\gamma^y\mu^z\sigma^z$ & $+$ & $+\gamma^x\mu^x\sigma^z$ \\
\end{tabular}
\end{ruledtabular}
\end{table}

\begin{table}[ht]
\caption{Bond-resolved DM textures of one unit cell (site labels of
Ref.~\cite{Romhanyi2015}; four of the eight orientations are reversed
relative to that reference, and primed labels denote the same sublattice
site in a neighboring cell, $A_1'=(2,0)$, $A_1''=(0,2)$, $B_1'=(1,2)$,
$B_2'=(2,1)$); each column is one symmetry mode. The
high-temperature structure admits only $\bm D^{(\perp)}$.}
\begin{ruledtabular}
\begin{tabular}{llccc}
bond $i\to j$ & sites & $\bm D^{(\perp)}/D'_\perp$ & $\bm D^{(s)}/D'_{\parallel,s}$ & $\bm D^{(ns)}/D'_{\parallel,ns}$ \\
\colrule
$(0,0)\to(1,0)$ & $A_1\to B_1$ & $(0,0,1)$ & $(-\tfrac{1}{\sqrt2},\tfrac{1}{\sqrt2},0)$ & $(-\tfrac{1}{\sqrt2},-\tfrac{1}{\sqrt2},0)$ \\
$(0,0)\to(0,1)$ & $A_1\to B_2$ & $(0,0,-1)$ & $(-\tfrac{1}{\sqrt2},\tfrac{1}{\sqrt2},0)$ & $(\tfrac{1}{\sqrt2},\tfrac{1}{\sqrt2},0)$ \\
$(0,1)\to(1,1)$ & $B_2\to A_2$ & $(0,0,-1)$ & $(-\tfrac{1}{\sqrt2},\tfrac{1}{\sqrt2},0)$ & $(-\tfrac{1}{\sqrt2},-\tfrac{1}{\sqrt2},0)$ \\
$(1,1)\to(1,0)$ & $A_2\to B_1$ & $(0,0,-1)$ & $(\tfrac{1}{\sqrt2},-\tfrac{1}{\sqrt2},0)$ & $(-\tfrac{1}{\sqrt2},-\tfrac{1}{\sqrt2},0)$ \\
$(1,0)\to(2,0)$ & $B_1\to A_1'$ & $(0,0,-1)$ & $(-\tfrac{1}{\sqrt2},-\tfrac{1}{\sqrt2},0)$ & $(-\tfrac{1}{\sqrt2},\tfrac{1}{\sqrt2},0)$ \\
$(0,1)\to(0,2)$ & $B_2\to A_1''$ & $(0,0,1)$ & $(\tfrac{1}{\sqrt2},\tfrac{1}{\sqrt2},0)$ & $(-\tfrac{1}{\sqrt2},\tfrac{1}{\sqrt2},0)$ \\
$(1,2)\to(1,1)$ & $B_1'\to A_2$ & $(0,0,1)$ & $(-\tfrac{1}{\sqrt2},-\tfrac{1}{\sqrt2},0)$ & $(\tfrac{1}{\sqrt2},-\tfrac{1}{\sqrt2},0)$ \\
$(1,1)\to(2,1)$ & $A_2\to B_2'$ & $(0,0,1)$ & $(-\tfrac{1}{\sqrt2},-\tfrac{1}{\sqrt2},0)$ & $(-\tfrac{1}{\sqrt2},\tfrac{1}{\sqrt2},0)$ \\
\colrule
\multicolumn{2}{l}{intra-dimer: $(0,0)\to(1,1)$ [$A_1\to A_2$]} & \multicolumn{3}{l}{$\bm D^{(D)}/D = (\tfrac{1}{\sqrt2},-\tfrac{1}{\sqrt2},0)$}\\
\multicolumn{2}{l}{\phantom{intra-dimer:} $(2,1)\to(1,2)$ [$B_2'\to B_1'$]} & \multicolumn{3}{l}{$\bm D^{(D)}/D = (-\tfrac{1}{\sqrt2},-\tfrac{1}{\sqrt2},0)$}\\
\end{tabular}
\end{ruledtabular}
\end{table}

\section*{Continuum theory and locked action}
The Majorana-matrix formulation~\cite{ThomsonSachdev,Feuerpfeil2026}
combines the spinons into $\mathcal F_i=\bigl(\begin{smallmatrix}f_{i\uparrow}&
-f_{i\downarrow}\\ f^\dagger_{i\downarrow}&f^\dagger_{i\uparrow}
\end{smallmatrix}\bigr)$ with $\mathcal F_i^\dagger=\sigma^y\mathcal
F_i^T\sigma^y$. SU(2) gauge transformations act on the left and spin
rotations $\Omega$ on the right, $\mathcal F\to\mathcal F\sigma^z
\Omega^T\sigma^z$. Near the $\pi$-flux nodes the field $\X$ carries Dirac
($\gamma^0=\rho^y$, $\gamma^x=\ii\rho^z$, $\gamma^y=\ii\rho^x$), valley
($\mu^a$), spin and gauge indices, $\Xb=\X^\dagger\gamma^0$, and the
emergent SO(5) has generators
$T^j=\{\mu^y,\sigma^a,\mu^x\sigma^a,\mu^z\sigma^a\}$ and vector
$\Gamma^j=\{\mu^x,\mu^z,\mu^y\sigma^a\}$. Projecting the lattice
projective symmetry group of Ref.~\cite{Feuerpfeil2026} onto the node
subspace gives $G_x:\X\to\rho^x\X(x,-y)$,
$\mathcal T:\X\to\rho^y\mu^y\X$ with $\ii\to-\ii$ and an additional
$\ii\sigma^y$ on the spin index, and $\sigma_{xy}:\X\to-\rho^z\mu^z
e^{-\ii\pi\rho^y/4}e^{\ii\pi\mu^y/4}\X(y,x)$. With DM each generator
carries its axial spin rotation as the right factor
$\sigma^z\Omega_O^T\sigma^z$: Eq.~(2) of the Letter and
$\hat\sigma_{x\bar y}=\sigma_{x\bar y}\otimes C_2^{[110]}$ for the buckled
structure. The high-temperature structure adds the mirror
$\hat\sigma_h=\one\otimes C_{2z}$, the pure glide $G_x\otimes C_{2y}$ and
the pure rotation $C_4\otimes R_z(+\pi/2)$, and with $D'_\perp$ alone the
full spin U(1)$_z$. The Higgs fields transform as in
the unlocked theory.

\section*{Independent construction in physical axes}
The classification is corroborated by an independent construction of
the continuum symmetry action and of the 36 gauge-singlet bilinears
directly in physical axes ($x\parallel[100]$, $y\parallel[010]$). It
yields the same inventory of six masses and 30 conserved-current
components, the vanishing of the ten SO(5)-adjoint scalars, and
Table~S2 under the identification $\sigma^x_{\rm label}=-S^x$. The
single invariant under the buckled locked group with $\mathcal T$ is
$J^x_x-J^y_y$, the Dresselhaus form of $\bar42m$. The bond-resolved DM
textures of Table~S4 project onto it alone, and $D'_\perp$ projects to zero. The locked space group alone forbids all six masses;
the pairs $\{\hat G_x,\hat\sigma_{xy}\}$, $\{\hat G_x,\hat\sigma_{x\bar y}\}$,
$\{\hat G_x,\hat S_4\}$, $\{\hat\sigma_{xy},\hat S_4\}$ and
$\{\hat\sigma_{x\bar y},\hat S_4\}$ each suffice, whereas time reversal
together with translations leaves both VBS masses allowed.

\section*{Comparison with published gauge-adjoint tables}
Under the continuum symmetry action of Tables~S1--S3, the transformations of the
gauge-adjoint bilinears $\Tr[\sigma^a\Xb M\X]$ agree entry by entry with
Tables~2 and~3 of Ref.~\cite{Feuerpfeil2026} ($G_x$, $\mathcal T$,
$\sigma_{xy}$; Pauli matrices written there as $\tau$)
and with Figs.~3 and 4 of Ref.~\cite{ThomsonSachdev} ($\mathcal T$, $P_x$,
$P_y$, $T_x$, $T_y$, $R_{\pi/2}$; the $R_{\pi/2}$ images agree up to the
sign convention of the representatives within each two-cycle), with one
exception. For the N\'eel-flavored currents
$\Tr[\sigma^a\Xb\mu^y\sigma^b\gamma^\mu\X]$, $\mu=0,x,y$,
Ref.~\cite{ThomsonSachdev} lists the time-reversal parities $(+,-,-)$,
whereas the present construction and Ref.~\cite{Feuerpfeil2026} give $(-,+,+)$.

The parities $(-,+,+)$ follow from two facts. Each of these bilinears
is Hermitian as written, because its kernel is a product of Hermitian
Pauli matrices, so no factor of $\ii$ enters. The antiunitary
$\mathcal T$ acts on a kernel $A$ as $A\to W^\dagger A^*W$, a
multiplicative map, and the Dirac, valley, spin and gauge matrices act on
different tensor factors. The N\'eel-flavored and valence-bond-flavored
currents with the same gauge and Dirac factors therefore differ in
time-reversal parity by the same factor as the N\'eel mass
$\Tr[\Xb\mu^y\X\sigma^b]$ and the valence-bond masses
$\Tr[\Xb\mu^{x,z}\X]$, which is $-1$: the N\'eel order parameter is odd
under time reversal and the valence-bond one even (Table~S1). Given the
published parities $(+,-,-)$ of the $\mu^x\gamma^\mu$ and
$\mu^z\gamma^\mu$ rows, the $\mu^y\sigma^b\gamma^\mu$ rows must be
$(-,+,+)$. The same conclusion follows within Ref.~\cite{ThomsonSachdev} from its
own table of $\Tr[\sigma^a\Xb T^j\X]$, whose time-reversal column
agrees with the present construction. The discrepancy leaves the gauge-singlet
classification of the Letter unchanged. It enters only the Yukawa
couplings of the companion paper that contain these currents, for which
the parities $(-,+,+)$ are used~\cite{companion}.

\end{document}